\documentclass[preprint]{elsarticle}

\usepackage{graphicx}
\usepackage{amsmath,amsthm}
\usepackage{amssymb,dsfont}
\usepackage{bm,color}
\usepackage{comment}
\usepackage{color}
\usepackage[T1]{fontenc}
\usepackage[utf8]{inputenc}
\usepackage[british]{babel}
\usepackage{algorithm,algorithmic}
\usepackage{xcolor}
\usepackage{subfigure}

\newcommand{\WRP}{\par\!\!\!\!\!\!\!\!\!\!}

\newtheorem{theorem}{Theorem}

\newtheorem{definition}{Definition}
\newtheorem{remark}[definition]{Remark}

\usepackage{etoolbox}

\definecolor{mediumtealblue}{rgb}{0.0, 0.33, 0.71}
\definecolor{forestgreen}{rgb}{0.0, 0.27, 0.13}
\definecolor{bittersweet}{rgb}{1.0, 0.44, 0.37}
\definecolor{azure}{rgb}{0.0, 0.5, 1.0}

\DeclareMathAlphabet{\mathonebb}{U}{bbold}{m}{n}

\DeclareMathOperator{\var}{Var}

\newcommand{\R}{{\mathbb R}}
\newcommand{\Z}{{\mathbb Z}}

\newcommand{\D}[2]{ \ensuremath{ \frac{\mathrm{d} #1 }{\mathrm{d} #2 } }}
\newcommand{\DP}[2]{ \ensuremath{ \frac{\partial #1 }{\partial #2 } } }

\newcommand{\proba}[1]{\ensuremath{{\mathbb P}\!\left[#1 \right]} }
\newcommand{\esp}[1]{\ensuremath{{\mathbb E}\!\left[#1 \right]} }

\renewcommand{\vec}[1]{\ensuremath{{\mathbf #1 } }}

\makeatletter
\newcommand*{\inlineequation}[2][]{%
  \begingroup
    \refstepcounter{equation}%
    \ifx\\#1\\%
    \else
      \label{#1}%
    \fi
    \relpenalty=10000 %
    \binoppenalty=10000 %
    \ensuremath{%
      #2%
    }%
    ~\@eqnnum
  \endgroup
}
\makeatother

\title{Push-forward method for piecewise deterministic biochemical simulations}

\author{Guilherme C.P. Innocentini} 
\address{Universidade Federal do ABC, Santo Andr\'e, Brazil}

\author{Arran Hodgkinson}
\address{Quantitative Biology and Medicine, University of Exeter, United Kingdom \\
Hodgkinson Laboratories Limited, Sheffield, United Kingdom}

\author{Fernando Antoneli}
\address{Escola Paulista de Medicina, Universidade Federal de S\~ao Paulo, S\~ao Paulo, Brazil}

\author{Arnaud Debussche}
\address{University of Rennes, CNRS, IRMAR - UMR 6625, F- 35000 Rennes, France}

\author{Ovidiu Radulescu}
\ead{ovidiu.radulescu@umontpellier.fr}
\address{University of Montpellier, CNRS, LPHI - UMR CNRS 5235, Montpellier, France}

\begin{document}

\begin{abstract}
A biochemical network can be simulated by a set of ordinary differential equations (ODE) under well stirred reactor conditions, for large numbers of molecules, and frequent reactions. 
This is no longer a robust representation when some molecular species are in small numbers and reactions changing them are infrequent. In this case, discrete stochastic events trigger changes of the smooth deterministic dynamics of the biochemical network. Piecewise-deterministic Markov processes (PDMP) are well adapted  for describing such situations. Although PDMP models are now well established in biology, these models remain computationally challenging. Previously we have introduced the push-forward method to compute how the probability measure is spread by the deterministic ODE flow of PDMPs, through the use of analytic expressions of the corresponding semigroup. In this paper we provide a more general simulation algorithm that works also for non-integrable systems. The method can be used for biochemical simulations with applications in fundamental biology, biotechnology and biocomputing. This work is an extended version of the work presented at the conference CMSB2019. 
\end{abstract}

\maketitle

\section{Introduction}

Stochastic simulation is a powerful tool in biology. Moreover, stochasticity is a general property of biochemical networks, having multiple origins. 
Noise is generated intrinsically by these molecular systems when neither the law of large numbers nor the averaging theorem can be applied, for instance when some of the constituent molecular species are present in small numbers and trigger relatively slow reactions \cite{crudu2009hybrid}. There are also extrinsic sources of noise, resulting from a fluctuating cellular environment. The extrinsic noise paradigm also applies to stochasticity resulting from manipulating biochemical networks in an artificial environment such as lab on a chip or simply in a titration device \cite{abate2018experimental}.  

Stochastic simulation is used in single cell experimental studies, where the amounts of mRNA \cite{thattai2004stochastic,raj2006stochastic,cai2006stochastic,tantale2016single} and protein \cite{elowitz2002stochastic,ferguson2012} products of a gene can be determined for each cell. By double- or multiple-fluorophore fluorescence techniques, products from several genes can be quantified simultaneously and one can have access to multivariate probability distributions of mRNA or proteins.
The stochastic dynamics of promoters and gene networks can have important consequences for fundamental biology \cite{eldar2010functional} but also for HIV \cite{razooky2015hardwired} and cancer research \cite{gupta2011stochastic}.
Predicting the probability distributions of biochemical networks' products is also instrumental for lab-on-a-chip applications, when one wants to  optimize and control the functioning of these networks. 
For these reasons we aim to develop effective methods for computing time-dependent distributions for stochastic models.
Our main objective is the reduction of computation time which is prerequisite for parameter scans and machine learning applications \cite{herbach2017inferring}.

The traditional stochastic simulation protocol uses the chemical master equation and the Gillespie algorithm. 
In such a protocol all the chemical reactions are simulated individually as discrete stochastic events.
Simulations on relevant experimental times have to cope with $10^6-10^9$ such events and generate $10^2-10^4$ samples in order to get statistically significant estimates of molecular species distributions. 
Altogether, these simulations are extremely costly in terms of execution time. 

An important reduction of the simulation time is obtained by noticing that the sources of noise can be localized in small parts of the system that behave discretely, or in  the discrete environmental variables in the case of intrinsic or extrinsic noise, respectively. The remaining, large part of the system consists of molecular species in large numbers that evolve continuously. 
This leads to piecewise-deterministic Markov process (PDMP) approximations of the biochemical dynamics, coupling discrete state Markov chain dynamics with continuous state ordinary differential equations (ODE) dynamics.  
The justification of the PDMP approximations of the chemical master equation can be found in \cite{crudu2009hybrid,crudu2011convergence}. 
Although simpler than the chemical master equation, direct simulation of the PDMP remains time consuming because the Markov chains can still have a very large number of states \cite{crudu2009hybrid}. 

In the CMSB2019 proceedings paper we have introduced new methods for simulating PDMPs for gene networks \cite{innocentini2019effective}. 
A gene network PDMP model can be simulated by numerical integration of ODEs satisfied by the mRNA and the protein species, coupled to the Markov chain describing the successive transitions
of the gene promoters 
\cite{zeiser2008simulation,crudu2009hybrid,riedler2013almost,lin2018efficient}. 
The simulation becomes particularly effective when analytic solutions of the ODEs are available \cite{innocentini2018time}.

Probability distributions of PDMP are solutions of the Liouville-master partial differential equations (PDEs).
Numerical integration of these PDEs is an interesting alternative to direct simulation, combining precision and speed for small models. Finite difference methods, however, are of limited use in this context as they can not cope with high dimensional models (for instance, extant gene networks applications are restricted to the dimension 2, corresponding to a single promoter, with or without self-regulation see \cite{innocentini2018time,kurasov2018stochastic}).

Another interesting method for computing time dependent distributions is the push-forward method.
For gene networks, this method has been first introduced in \cite{innocentini2016protein}
and further adapted for continuous mRNA variables in \cite{innocentini2018time}.
It is based on the idea to compute the probability distribution of gene products 
as the push-forward measure of the semigroup defined by the ODEs. This method is an approximation, as one has to consider that the discrete PDMP variables are piecewise constant on a deterministic time partition. The transition rates between promoter states were computed using a mean field approximation in \cite{innocentini2018time}.
In the CMSB2019 proceeding paper, the mean field approximation was replaced by the next order approximation taking into account the second moment of the protein distribution \cite{innocentini2019effective}.  In this paper we extend the method to general PDMP models, thus covering all biochemical networks, with both intrinsic and extrinsic  noise simulation protocols.  


\section{Models}

\subsection{ODE models of biochemical networks}

A biochemical network is defined by a set of chemical reactions and a set of chemical 
species. The amounts of different chemical species form a vector \linebreak  $x = (x_1,x_2,\ldots,x_N) \in \R^N_+$, where $x_i$ is the concentration of the species $i$. Each reaction is characterized by a stoichiometric vector $\nu_i \in \Z^N$ and a reaction rate function $R_i: \R^N_+ \to \R$.
When all chemical species are present in large numbers, the biochemical network is well described by a systems of ODEs
\begin{equation} \label{eq:ODE_MODEL}
\D{x}{t} = \sum_{i=1}^r \nu_i R_i(x),
\end{equation}

For example, a simple gene transcription model can be obtained when a large number, let's say $G$, of a specific gene is present in the cell and, moreover, the gene promoter has just two states: active and inactive. In this simple model the gene product is the mRNA concentration. The vector of species concentration is given by: $x_1=P$ (inactive promoter concentration), $x_2=P^*$ (active promoter concentration) and $x_3=X$ (mRNA concentration). The concentration of inactive promoters is measured by the ratio between the number of inactive promoters ($G_I$) and the cell volume $V_{\mathrm{cell}}$, so $P=G_I/V_{\mathrm{cell}}$; the same holds for the concentrations of active promoters and mRNA, $P=G_A/V_{\mathrm{cell}}$, where $G_A$ is the number of active promoters.
The reactions are $P \to P^*$, $P^* \to P$, $P^* \to P^* + X$, $X \to \varnothing$. Thus, the stoichiometric vectors are $\nu_1 = (-1,1,0)$, $\nu_2 = (1,-1,0)$, $\nu_3 = (0,0,1)$, $\nu_4 = (0,0,-1)$ and reaction functions $R_1 = k_1 P$, $R_2=k_2 P^*$, $R_3 =k_3 P^*$, $R_4 = k_4 X$, where the reaction rate constants $k_i,\, 1\leq i \leq 4$, have dimension $[1/\mathrm{time}]$.
For large copy numbers of $P$, $P^*$ and $X$ the simple transcription model reads:
\begin{equation} \label{eq:STM}
\D{P}{t} = -k_1 P + k_2 P^*, \;
\D{P^*}{t} = k_1 P - k_2 P^*, \;
\D{X}{t} = k_3 P^* - k_4 X \,.
\end{equation}
Note that $\D{P}{t}+\D{P^ *}{t}=0$ and so one of the first two equations can be discarded.

We must emphasize that gene transcription is only one of the many possible examples of our formalism. The formalism defined by \eqref{eq:ODE_MODEL} covers all biochemical network models and have large applicability in cell physiology.

\subsection{PDMP models}\label{s2.2}

When some, but not all, species are present in low numbers and there are also slow reactions, a stochastic representation, such as a piecewise deterministic Markov process (PDMP), is more appropriate than the deterministic equations \eqref{eq:ODE_MODEL}.

For instance, in the simple transcription model \eqref{eq:STM}, assume that there is only one copy of the gene, meaning  $G=G_A+G_I=1$.
In this case, it does not make sense to describe the time evolution of promoter variables by differential equations anymore. In this scenario, the promoter variables are discrete
numbers.  As the copy number of the gene is one, is more suitable to replace the concentration $P$ by the discrete variable $G_A$
assuming values $G_A=1$ or $G_A=0$. The same holds for $P^*$, 
that is replaced by $G_I \in \{0,1\}$.
Moreover, since $G_A+G_I=1$ one of these variables can be discarded.

The switching between the two discrete states of $G_A$ can be described by a continuous time Markov chain with transition rates $k_1$ and $k_2$. That is, if $G_A(0)=1$, then $G_A(t)=1$ for $0 \leq  t < T_1$ where $T_1$ is a random time such that $\proba{T_1 > t} = \exp(-k_2 t)$ and $G_A(t)=0$ for $T_1 \leq t < T_1 + T_2$ where $T_2$ is a random time such that $\proba{T_2 > t} = \exp(-k_1 t)$.

Moreover, suppose that the switching constants, $k_1$ and $k_2$, are small compared to $k_4$, and also that $k_3 /k_4 \gg 1$. Thus, the variable $X$ has a switching behavior alternating accumulation and degradation periods when $G_A=1$ and $G_A=0$, respectively, each one described by ODEs:
\begin{equation} \label{eq:HYBRID}
\D{X}{t} = \left\{
\begin{array}{lll}
k_3/V_{\mathrm{cell}} - k_4 X & \text{when} &G_A=1 \\
-k_4 X & \text{when} & G_A=0 \\
\end{array}
\right. ,
\end{equation}
where $V_{\mathrm{cell}}$ is the cell volume.

Using methods from \cite{crudu2011convergence} it is possible to show that the ``hybrid system'' $(G_A/V_{\mathrm{cell}},X)$, with $X$ given by \eqref{eq:HYBRID}, converges to \eqref{eq:STM}, as $G \to\infty$.

The PDMP formalism can be extended to a rather general class of biochemical network models as follows:

A PDMP biochemical model is a stochastic process $\vec{\zeta}_t$ having states in a set of the form $E = \R^{N}_+ \times {\mathcal S}$, where ${\mathcal S}$ is a finite set encoding discrete states of the biochemical model, and $\vec{\zeta}_t = (\vec{x}_t,s_t)$, where $\vec{x}_t$ is a vector in $\R^N_+$ whose components $x^{i}_t$ ($1 \leq i \leq N$) encode the dynamics of $i$ continuous biochemical species, and $s_t$ describes the jump Markov process between the discrete states. The PDMP $\vec{\zeta}_t = (\vec{x}_t,s_t)$ is determined by three characteristics:
\begin{description}
\item[1)]
For all fixed $s \in {\mathcal S}$, a vector field $\vec{V}_{s} : \R^{N}_+ \to \R^{N} $ determining a unique global flow $\vec{\Phi}_{s}(t,\vec{x})$ in $\R^{N}_+$, such that, for $t>0$,
\begin{equation}
\D{\vec{\Phi}_{s}(t,\vec{x})}{t} =  
\vec{V}_{s}(\vec{\Phi}_{s}(t,\vec{x})), \;
\vec{\Phi}_{s}(0,\vec{x})=\vec{x}. \label{flow}
\end{equation}


The flow $\vec{\Phi}_{s}(t,\vec{x})$ represents a one parameter semigroup fulfilling the properties
\begin{description}
\item[(i)]
$\vec{\Phi}_{s}(0,\vec{x}_0) = \vec{x}_0$,
\item[(ii)]
$\vec{\Phi}_{s}(t+t',\vec{x}_0) = 
\vec{\Phi}_{s}(t',\vec{\Phi}_{s}(t,\vec{x}_0))$.
\end{description}

\item[2)]
A transition rate matrix $\vec{H} : \R^{N}_+ \to  M_{ N_s \times N_s} (\R)$, such that $H_{r,s}(\vec{x})$ is the $(r,s)$ element of the matrix $\vec{H}$, $N_s = \# {\mathcal S}$ is the number of states. If $(s\neq r)$, $H_{r,s}(\vec{x})$  is the rate of probability to jump to the state $r$ from the state $s$.
Furthermore, $H_{s,s}(\vec{x}) =  - \sum_{r \neq s} H_{r,s}(\vec{x})$ for all $s \in {\mathcal S}$ and for all $\vec{x}\in R^{N}$.

\item[3)]
A jump rate $\lambda : E \to \R_+$. The jump rate can be obtained from the transition rate matrix
\begin{equation}\label{transition}
\lambda(\vec{x},s) = \sum_{r\neq s} H_{r,s}(\vec{x}) = - H_{s,s}(\vec{x}).
\end{equation}
\end{description}

From these characteristics, right-continuous sample paths $\{\vec{x}_t : t >0 \}$ starting at $\vec{\zeta}_0=(\vec{x}_0,s_0) \in E$ can be constructed as follows. Define
\begin{equation}
\vec{x}_t (\omega) := \vec{\Phi}_{s_0}(t,\vec{x}_0)
\text{ for } 0 \leq t \leq T_1(\omega),
\end{equation}
where $T_1(\omega)$ is a realisation of the first jump time of $s_t$, with the distribution
\begin{equation}
F(t) = \proba{T_1 > t} = \exp \left(-\int_{0}^{t} \lambda(\vec{\Phi}_{s_0}(u,\vec{x}_0),s_0 )\,du \right), \; t>0, \label{next}
\end{equation}
and $\omega$ is the element of the probability space for which the particular realisation of the process is given.
The pre-jump state is $\vec{\zeta}_{T_1^-(\omega)} (\omega) = (\vec{\Phi}_{s_0}(T_1(\omega),\vec{x}_0), s_0 )$ and the post-jump state
is  $\vec{\zeta}_{T_1(\omega)} (\omega) = 
(\vec{\Phi}_{s_0}(T_1(\omega),\vec{x}_0), s )$, where $s$ has the distribution
\begin{equation}
\proba{s=r} = \frac{H_{r,s_0} (\vec{\Phi}_{s_0}(T_1(\omega),\vec{x}_0),s_0) }{ \lambda(\vec{\Phi}_{s_0}(T_1(\omega),\vec{x}_0),s_0)}, \text{ for all }
r \neq s_0.
\end{equation}
We then restart the process  $\vec{\zeta}_{T_1(\omega)}$ and recursively apply the same procedure at jump times $T_2(\omega)$, etc..

Note that between each two consecutive jumps $\vec{x}_t$ follow deterministic ODE dynamics defined by the vector field $\vec{V}_{s}$. At the jumps, the values $\vec{x}_t$  are continuous. More general definitions of PDMPs include jumps in the continuous variables but will not be discussed here. 
  

We define multivariate probability density functions $p_{s}(t,\vec{x})$.
These functions satisfy the Liouville-master equation which is a system of partial differential equations:
\begin{equation}
\DP{p_{s}(t,\vec{x})}{t} =
- \nabla_{\vec{x}} . (\vec{V}_{s}(\vec{x}) p_{s}(t,\vec{x}) )
+  \sum_{s'} H_{s,s'}(\vec{x}) 
p_{s'}(t,\vec{x}). 
\label{master-liouville}
\end{equation}

This general formalism covers all biochemical models that have discrete variables. As salient examples we can cite gene networks \cite{innocentini2018time}, ion channels dynamics for neuron networks  \cite{anderson2015ionchannel}, lab on chip  biochemical devices \cite{abate2018experimental}.

\subsection{PDMP models of ON/OFF gene networks}

A particular example of PDMP biochemical model is represented by the gene networks. Because these models are extensively used in systems biology, we provide their details in this subsection. 

Two state ON/OFF gene networks generalize the simple two state single gene transcription model, by considering more interacting genes and by using separate variables for the two gene products: mRNA and protein. 
 
To each gene we associate a discrete variable $\sigma_i$ with two possible values $\sigma_i=0$ for a non-productive state (OFF)  and $\sigma_i=1$ for a productive state (ON).
Furthermore, a gene $i$ produces proteins and mRNAs in the amounts $y_i$ and $r_i$, respectively.
A gene network state is described by the $N-$dimensional vector $x_t=(r_1(t),y_1(t),r_2(t),y_2(t),\ldots,r_{N/2}(t),y_{N/2}(t))$.
The gene products amounts satisfy ODEs:
\begin{eqnarray}
\D{y_i}{t} & = & b_i r_i - a_i y_i, \notag \\
\D{r_i}{t} & = & k_i (\sigma_i) - \rho_i r_i \label{standard1}
\end{eqnarray} 
The coupling between genes is modeled at the level of discrete state transitions. 
The elements of the matrix $\vec{H}$ are functions of products from various genes. 
For instance, if a gene $i$ inhibits a gene $j$ the transition rate from ON to OFF of the gene $j$ is an increasing function of the protein concentration $y_i$. 

As a first example that we denote as model $M_1$, let us consider a two genes network; the expression of the first gene being constitutive and the expression of the second gene being activated by the first. We consider that the transcription activation rate of the second gene is proportional to the concentration of the first protein $f_2 y_1$. All the other rates are constant $f_1$, $h_1$, $h_2$, representing the transcription activation rate of the first gene, and the transcription inactivation rates of gene one and gene two, respectively.
For simplicity, we consider that the two genes have identical protein and mRNA parameters $b_1 = b_2 = b$, $a_1 = a_2 = a$, $\rho_1 = \rho_2 = \rho$.
We further consider that $k_i = k_0$ if the gene $i$ is OFF and $k_i = k_1 > k_0$ if the gene $i$ is ON.

The gene network has four discrete states,  in order $(0,0)$, $(1,0)$, $(0,1)$, and $(1,1)$.
Then, the transition rate matrix for the model $M_1$ is
\begin{equation}
\vec{H}(y_1,y_2)=
\begin{bmatrix}
-(f_1 + f_2y_1) & h_1 & h_2 & 0 \\
f_1 & -(h_1+f_2y_1) & 0 & h_2  \\
f_2y_1 & 0 & -(f_1 + h_2) & h_1 \\
0 & f_2y_1  & f_1 & -(h_1+h_2)
\end{bmatrix}.
\end{equation}
The Liouville-master equation for the model $M_1$ reads
\begin{equation} \begin{aligned}
\DP{p_1}{t} = &
-\DP{[(b r_1 - a y_1) p_1]}{y_1} - \DP{ [(k_0 - \rho r_1) p_1]}{r_1}
-\DP{[(b r_2 - a y_2) p_1]}{y_2} - \DP{ [(k_0 - \rho r_2) p_1]}{r_2} \\
& + h_2 p_3 + h_1 p_2 - (f_1 + f_2y_1)  p_1, \\[2mm]
\DP{p_2}{t} = &
-\DP{[(b r_1 - a y_1) p_2]}{y_1} - \DP{ [(k_1 - \rho r_1) p_2]}{r_1}
-\DP{[(b r_2 - a y_2) p_2]}{y_2} - \DP{ [(k_0 - \rho r_2) p_2]}{r_2} \\
& + f_1 p_1 + h_2 p_4 - (h_1 + f_2y_1)  p_2, \\[2mm]
\DP{p_3}{t} = &
-\DP{[(b r_1 - a y_1) p_3]}{y_1} - \DP{ [(k_0 - \rho r_1) p_3]}{r_1}
-\DP{[(b r_2 - a y_2) p_3]}{y_2} - \DP{ [(k_1 - \rho r_2) p_3]}{r_2} \\
& + h_1 p_4 + f_2 y_1 p_1 - (h_2 + f_1)  p_3, \\[2mm]
\DP{p_4}{t} = &
-\DP{[(b r_1 - a y_1) p_4]}{y_1} - \DP{ [(k_1 - \rho r_1) p_4]}{r_1}
-\DP{[(b r_2 - a y_2) p_4]}{y_2} - \DP{ [(k_1 - \rho r_2) p_4]}{r_2} \\
& + f_1 p_3+ f_2 y_1 p_2 - (h_1 + h_2)  p_4.
\label{liouville-master-twogene1}
\end{aligned} \end{equation}
The model $M_2$ differs from the model $M_1$ by the form of the activation function. 
Instead of a linear transcription rate $f_2 y_1$ we use a Michaelis-Menten model $f_2 y_1/(K_1 + y_1)$. 
This model is more realistic as it takes into account that the protein $p^1$ has to attach to specific promoter sites which become saturated when the concentration of this protein is high.

The transition rate matrix for the model $M_2$ is
\begin{equation} \setlength{\arraycolsep}{2.5pt} 
\vec{H}(y_1,y_2)=
\begin{bmatrix}
-(f_1 + \frac{f_2y_1}{K_1+y_1}) & h_1 & h_2 & 0 \\
f_1 & -(h_1+\frac{f_2y_1}{K_1+y_1}) & 0 & h_2  \\
\frac{f_2y_1}{K_1+y_1} & 0 & -(f_1 + h_2) & h_1 \\
0 & \frac{f_2y_1}{(K_1+y_1)}  & f_1 & -(h_1+h_2)
\end{bmatrix}.
\end{equation}
The Liouville-master equation for the model $M_2$ reads
\begin{align*}
\DP{p_1}{t} =&
-\DP{[(b r_1 - a y_1) p_1]}{y_1} - \DP{ [(k_0 - \rho r_1) p_1]}{r_1}
-\DP{[(b r_2 - a y_2) p_1]}{y_2} - \DP{ [(k_0 - \rho r_2) p_1]}{r_2} \\
& + h_2 p_3 + h_1 p_2 - (f_1 + f_2 y_1 / (K_1 + y_1))  p_1, \\[2mm]
\DP{p_2}{t} = &
-\DP{[(b r_1 - a y_1) p_2]}{y_1} - \DP{ [(k_1 - \rho r_1) p_2]}{r_1}
-\DP{[(b r_2 - a y_2) p_2]}{y_2} - \DP{ [(k_0 - \rho r_2) p_2]}{r_2} \\
& + f_1 p_1 + h_2 p_4 - (h_1 + f_2 y_1 / (K_1 + y_1))  p_2, \\[2mm]
\DP{p_3}{t} = &
-\DP{[(b r_1 - a y_1) p_3]}{y_1} - \DP{ [(k_0 - \rho r_1) p_3]}{r_1}
-\DP{[(b r_2 - a y_2) p_3]}{y_2} - \DP{ [(k_1 - \rho r_2) p_3]}{r_2} \\
& + h_1 p_4 + f_2 y_1 / (K_1 + y_1) p_1 - (h_2 + f_1)  p_3, \\[2mm]
\DP{p_4}{t} = &
-\DP{[(b r_1 - a y_1) p_4]}{y_1} - \DP{ [(k_1 - \rho r_1) p_4]}{r_1}
-\DP{[(b r_2 - a y_2) p_4]}{y_2} - \DP{ [(k_1 - \rho r_2) p_4]}{r_2} \\
& + f_1 p_3 + f_2 y_1 / (K_1 + y_1) p_2 - (h_1 + h_2)  p_4.
\label{liouville-master-twogene2}
\end{align*}



\section{Simulation methods}

\subsection{Monte-Carlo method}
The Monte-Carlo method utilizes the direct simulation of the PDMP based on the iteration of the following equations:
\begin{eqnarray}
\D{\vec{x}}{t}& = & \vec{V}_{s_0}(\vec{x}(t)), \notag \\
\D{ F}{t} &=& - \lambda(\vec{x},s_0) F, \label{standard2} 
\end{eqnarray}
for $t\in [0,T_1)$ with initial conditions $\vec{x}(0)=\vec{x}_0$, $F(0)=0$, and stopping condition $F(T_1)=U$, where $U$ is random variable, uniform in the range $[0,1)$, followed by the choice of the next discrete state $s_1$ by using the matrix $\vec{H}(\vec{x}(T_1))$.

A large number $N_{mc}$ of sample traces is generated and the values of $\vec{x}_t$ are stored at selected times. Multivariate time dependent probability distributions are then estimated from this data. 

The complete algorithm is presented in the Algorithms~\ref{alg:ns} and \ref{alg:mc}.

\begin{algorithm}[!ht]
  \begin{algorithmic}[1]
    \caption{\label{alg:ns}$\operatorname{NextState}$}
    \REQUIRE 1.~$x$ initial value continuous variable, 2.~$s$ initial discrete state,  3.~$t_0$ initial time, 4.~$t_{max}$ maximal time 
    \smallskip
    \ENSURE 1.~$(x_0,x_1,\ldots,x_n)$, 2.~$(t_0,t_1,\ldots,t_n)$ 3.~$s$ new discrete state 
    \smallskip
    \STATE{$x_0:=x$;\, $s_0:=s$}
    \STATE{$U := randunif([0,1))$}
    \STATE{solve $\D{x}{t}= V_{s_0}(x),\, \D{F}{t} = -\lambda(x,s_0) F$ with initial data
    $x(t_0)=x_0$,$F(t_0)=1$, until $F(t)=U$ or $t=t_{max}$, return 
    $x_0=x(t_0),\ldots,x_n=x(t_n)$, where
    $t_0 < t_1 < \ldots < t_n$, $t_n = \min (F^{-1}(U), t_{max})$.}
    \STATE{$V : = randunif([0,1))$}
    \STATE{$\mathit{CUM} :=0$;\, $I:=0$}
    \WHILE{$\mathit{CUM} < V$}
    \STATE{$I:=I+1$}
    \STATE{$\mathit{CUM}:=\mathit{CUM}+H_{I,s_0}(x)/\sum_{J\neq s_0 } H_{J,s_0}(x) $}
    \ENDWHILE
    \RETURN ($(x_0,x_1,\ldots,x_n),(t_0,t_1,\ldots,t_n),I$)
  \end{algorithmic}
\end{algorithm}

\begin{algorithm}[!ht]
  \begin{algorithmic}[1]
    \caption{\label{alg:mc}$\operatorname{PDMPmontecarlo}$}
    \REQUIRE 1.~$p_0(x,s)$, initial distribution of $(x_0,s_0)$. 2.~$\tau$ time step for computing the time-dependent distribution. 3.~$\delta x$ bin size in one direction.  
    \smallskip
    \ENSURE $n_t$ vectors $H_0,\ldots,H_{nt}$, \par representing the distribution at times $0,\tau,\ldots,n_t \tau$
    \smallskip
    \FOR{$i:=1$ \TO $n_t$}
    \STATE{$H_i :=$ $n_x^M$ dimensional null vector  }
    \ENDFOR
    \FOR{$m:=1$ \TO $N_{mc}$}
    \STATE{Draw $(x_0,s_0)$ from the distribution $p_0(x,s)$}
    \STATE{$x:=x_0$,$s:=s_0$}
    \STATE{$t:=t_0$;  $i:=0$}
    \WHILE{ $t < n_t \tau$}
    \STATE{$((x_0,\ldots,x_n),(t_0,\ldots,t_n),s_{next}):=\text{NextState}(x,s,t,t_M)$}
    \FOR{$j := 1$ \TO $n$}
    \IF{$t_j \leq i\tau < t_{j+1}$}
    \STATE{$X_i := (x_j + x_{j+1})/2$}
    \STATE{increment by $1/((\delta x)^N N_{mc})$, \par the bin corresponding to $X_i$ in the histogram $H_i$}
    \STATE{$i:=i+1$}
    \ENDIF
    \ENDFOR
    \STATE{$x:=x_n$}
    \STATE{$s:=s_{next}$}
    \STATE{$t:=t_n$}
    \STATE{$nr:=nr+1$}
    \ENDWHILE
    \ENDFOR
    \RETURN{$(H_0,\ldots,H_{n_t})$}
  \end{algorithmic}
\end{algorithm}

The direct simulation of PDMPs needs the solutions of Eqs. \eqref{standard2} which can be obtained by numerical integration. 
This is not always computationally easy.
Problems may arise for fast switching promoters when the ODEs have to be integrated many times on small intervals between successive jumps.
Alternatively, the numerical integration of the ODEs can be replaced by analytic solutions or quadrature. 
Analytic expressions are always available for the gene network flow \eqref{standard1} and read
\[
\begin{split}
\Phi_i^y(t,y_0,r_0) & = y_0 \exp ( - a_i t) \\
& + b_i \left[\left(r_0 - \frac{k_i(\sigma_i)}{\rho_i}\right)\frac{\exp (-\rho_i t)-1}{a_i-\rho_i}+ \frac{k_i(\sigma_i)}{\rho_i}\frac{1-\exp(-a_i t)}{a_i} \right], \\[2mm]
\Phi_i^r(t,y_0,r_0) & = (r_0 - k_i/\rho_i)\exp ( - \rho_i t) + k_i/\rho_i.
\end{split}
\]

Let us consider the following general expression of the jump intensity function 
\[
\lambda(\vec{x},s) = c_0(s) + \sum_{i}^N c_i (s) y_i + \sum_{i}^N d_i (s) f_i(y_i),
\]
where $f_i$ are non-linear functions, for instance Michaelis-Menten function
\[
f_i(p^i) = \frac{p^i}{K_i + p^i}
\]
or Hill functions 
\[
f_i(y_i) = \frac{(y_i)^{n_i}}{K_i^{n_i} + (y_i)^{n_i}}.
\]
If  $d_i=0$ for all $1\leq i \leq N$, the cumulative distribution function of the waiting time $T_1$ can be solved analytically \cite{innocentini2018time}, otherwise it can be obtained by quadrature. For example, for the model $M_2$ one has
\[
\begin{split}
\lambda(\vec{x},s) =  
& \left(f_1 + f_2 \frac{y_1}{K_1 + y_1}\right)\delta_{s,1} +
\left(h_1 + f_2 \frac{y_1}{K_1 + y_1}\right) \delta_{s,2} \\[1mm]
& + (h_2 + f_1) \delta_{s,3} + (h_2 + h_1) \delta_{s,4},
\end{split}
\]
where $\delta_{i,j}$ is Kronecker's delta.
In this case, the waiting time $T_1$ is obtained as the unique solution of the equation
\begin{equation}
\begin{split}
-\log (U) = & \bigg[(f_1 + f_2)T_1 +f_2 \int_0^{T_1} \frac{1}{K_1 + \Phi_1^y(t',y_0,r_0)}\, dt' \bigg] \delta_{s_0,1} \\
& + \bigg[(h_1 + f_2)T_1 +  f_2\int_0^{T_1} \frac{1}{K_1 + \Phi_1^y(t',y_0,r_0)} \, dt' \bigg] \delta_{s_0,2} \\[1mm]
& + (h_2 + f_1)T_1 \delta_{s_0,3} + (h_2 + h_1)T_1 \delta_{s_0,4},
\end{split}\label{T1}
\end{equation}
where $U$ is a random variable, uniformly distributed in the range $[0,1)$.
In our implementation of the algorithm we solve \eqref{T1} numerically, using the bisection method.


\subsection{Finite difference Liouville master equation method}

The finite difference Liouville master equation method for a given number of genes, $N_g$, uses a discrete approximation of the domain to compute the numerical solution across time for a given higher-dimensional system \eqref{liouville-master-twogene1}, with initial conditions given by $p_1(0,y_1,r_1,\ldots):=\delta(y_1,r_1,\ldots)$ and $p_2(0,r_1,y_1,\dots):=\ldots:=p_{2N_g}(0,y_1,r_1,\ldots):=0$ where $\delta(\cdot)$ is the Dirac delta function. We compute the solution for $2N_g$ distributions, since each gene has both an ON and OFF state, respectively.

In order to achieve the simulation of the equations given by system \eqref{liouville-master-twogene1}, we begin by discretising each of the domains into $n_\eta$ intervals, whose centres are given by $\{\eta_1,\ldots,\eta_{n_\eta}\}\in\vec{y}$ meeting the condition that $\eta_1<\eta_2<\ldots<\eta_{n_\eta}$ and where, likewise, an arbitrary discrete value in $y_i$ would be denoted $\eta_{i,j}$. Each $\eta{i,j}$ then represents a unique, discrete, $j^{\text{th}}$ position in the protein abundance domain spanned by $y_i$. Likewise, mRNA would have an analogous discretisation given by all $\rho_{i,j},\,\forall j\in\{1,\ldots,n_\rho\}$. Time is then similarly discretised by a time-step $\tau$ into $n_t+1$ temporal locations given by $\{0,\tau,\ldots,\tau n_t\}$. The task then becomes the computation of the solution at each of these discrete locations in the domain, such that the sink and source terms may be trivially calculated but where the derivative terms warrant further explanation.

The solution to the advection equation under a uniform coefficient, $\chi$, is given by a translation in the relevant domain. To achieve this, whilst also maintaining the stability of the system, as a whole, we implement a simultaneous forwards- and backwards-difference discrete operator scheme. This means that for an arbitrary probability density function, $p_k(t,y_i,\ldots)$, we write the discrete partial derivative operator as
\begin{equation}
    \chi\frac{\partial}{\partial y_i}p_k(t,\eta_{i,j},\ldots) =
        \left\{\begin{array}{c l}
            |\chi|\dfrac{p_k(t,\eta_{i,j-1},\ldots) - p_k(t,\eta_{i,j},\ldots)}{\eta_{i,j} - \eta_{i,j-1}} ~\
            &   \text{if } \chi>0   \\
            |\chi|\dfrac{p_k(t,\eta_{i,j+1},\ldots)-p_k(t,\eta_{i,j},\ldots)}{\eta_{i,j+1} - \eta_{i,j}} ~\
            &   \text{otherwise.}
        \end{array}\right.
    \label{eq:fb_diff}
\end{equation}
This may be evaluated as such for each term within the system of equations, given by \eqref{liouville-master-twogene1}, and guarantees the probability balance of the system, as a whole.

Beyond the computation of the equation's solutions at a single time-step the solutions must be computed robustly across time. We therefore employ a McCormack predictor-corrector scheme \cite{maccormack2001advection}, given explicitly for any $i^{\text{th}}$ time-step by
\begin{equation} 
\begin{aligned}
    p_{k,i+1}' =\, & p_{k,i} + \tau F_k(p_{1,i},\ldots,p_{N_g,i}) \\
    p_{k,i+1} =\, & \frac{1}{2}(p_{k,i} + p_{k,i+1}') + \frac{1}{2} \tau F_k(p_{1,i+1}',\ldots,p_{N_g,i+1}') \\&&
    \forall k\in\{1,\ldots,2N_g\},
\end{aligned} \label{eq:maccormack} 
\end{equation}
where $p_{k,i} = p_k(t_i,y_1,r_1,\ldots)$, $F_k(p_1,\ldots,p_{2N_g}) = \partial p_k/\partial t$ and $p_{k,i}'$ is the discrete nomenclature for a prediction of the solution for $p_k$ at time $t_i$.

The method for solving the problem, in totality, is then given by Algorithm \ref{alg:pde} where, again, the partial derivative terms on the right-hand side of each equation are evaluated using \eqref{eq:fb_diff}. Algorithm \ref{alg:pde} is a direct implementation of \eqref{eq:maccormack} as an algorithm with the concurrent calculation of the distributions $P_t,\,\forall t\in\{0,\tau,\ldots,n_t\tau\}$ from the individual distributions $p_{k,i},\, \forall k\in\{1,\ldots,2N_g\},\, i\in\{0,\ldots,n_t\}$. For each time-step, we solve the right-hand side of the equation using the predictor corrector scheme, update the value of each distribution, and calculate the total distribution, $P_t$.

\begin{algorithm}[!ht]
  \begin{algorithmic}[1]
    \caption{\label{alg:pde} $\operatorname{PDMPLiouvillemaster}$}
    \REQUIRE 1. $p_{1,0}$, $p_{2,0}$, $p_{3,0}$, $p_{4,0}$ initial distributions in $(y_1,r_1,\ldots,y_{N_g},r_{N_g})$, where $p_{i,t}:=p_i(t,y_1,r_1,\ldots,y_{N_g},r_{N_g}),$ $\forall i\in\{1,\ldots,2N_g\}$. 2. $\tau$ time step for computing the time-dependent distribution. 3. $\delta y,\,\delta r$ bin size for the protein and mRNA distributions, respectively.
    \smallskip
    \ENSURE $n_t+1$ distributions $P_0,\,\ldots,\,P_{n_t}$ representing the sums of the individual distributions in $(y_1,r_1,y_2,r_2)$ at times $0,\,\tau,\,\ldots,\,n_t\tau$, respectively
    \smallskip
    \FOR{$i:=1$ \TO $n_t$}
    \FOR{$j:=1$ \TO $2N_g$}
        \STATE{Use $p_{k,i-1}$ and discretised right hand side of PDE in \eqref{liouville-master-twogene1} to compute the temporal gradient, $\partial_t p_{j,i-1}$}
        \STATE{Compute predicted solution at next time step as $p_{j,i}'=p_{j,i-1} + \tau\partial_t p_{j,i-1}$}
    \ENDFOR
    \STATE{Set $P_i=0$}
    \FOR{$j:=1$ \TO $2N_g$}
        \STATE{Use prediction $p_{k,i}'$ and discretised right hand side of PDE in \ref{liouville-master-twogene1} to compute the temporal gradient, $\partial_t p_{j,i}'$}
        \STATE{Compute corrected solution at next time step as \\$p_{j,i}=\frac{1}{2}(p_{j,i-1} + p_{j,i}') + \frac{1}{2}\tau\partial_t p_{j,i}'$}
        \STATE{Set $P_i = P_i + p_{j,i}$}
    \ENDFOR
    \ENDFOR
    \RETURN{$(P_0,\,\ldots,\,P_{n_t})$}
  \end{algorithmic}
\end{algorithm}

\subsection{Push-forward method}

\subsubsection{General algorithm}
This method allows one to compute the multivariate probability distribution of the continuous variable $\vec{x}$ at a time $\tau$ given the probability distribution of $(\vec{x},s)$ at time $0$.


\begin{algorithm}[!ht]
  \begin{algorithmic}[1]
    \caption{\label{alg:pf}$\operatorname{PDMPpushforward}$}
    \REQUIRE 1.~$p_0(x,s)$, initial distribution of $(x_0,s_0)$. 2.~$\tau$ time step for computing the time-dependent distribution 
\smallskip
    \ENSURE $n_t$ vectors $H_0,\ldots,H_{nt}$, \par representing the distribution at times $0,\tau,\ldots,n_t \tau$
    \smallskip
    \STATE{compute $H_0,P0$ from initial distribution}
     \FOR{$i:=1$ \TO $n_t$}
     \FOR{$j:=1$ \TO $n_x^N$}
     \FOR{$s0:=1$ \TO $N_s$}
          \STATE{solve $\D{\vec{x0}}{t} = V_{s_0}(\vec{x0}), \D{\vec{\Pi0}}{t} = H(\vec{x0}) \vec{\Pi0} $ with initial conditions $\vec{x0}(0)=C_j, \Pi0(0)=\vec{I}$ from $t=0$ to $t=\tau_1$}
     \FOR{$s1:=1$ \TO $N_s$}
          \STATE{solve $\D{\vec{x1}}{t} = V_{s_1}(\vec{x1}), \D{\vec{\Pi1}}{t} = H(\vec{x1}) \vec{\Pi1} $ with initial conditions $x1(\tau_1)=x0(\tau_1),\Pi1(0)=\vec{I}$ from $t=\tau_1$ to $t=\tau_2$}
     \FOR{$s2:=1$ \TO $N_s$}
          \STATE{solve $\D{\vec{x2}}{t} = V_{s_2}(\vec{x2}), \D{\vec{\Pi2}}{t} = H(\vec{x2}) \vec{\Pi2} $ with initial conditions $x2(\tau_2)=x1(\tau_2), \Pi2(0)=\vec{I}$ from $t=\tau_2$ to $t=\tau_3$}
     \FOR{$s3:=1$ \TO $N_s$}
     \STATE{solve $\D{\vec{x3}}{t} = V_{s_3}(\vec{x3})$ with initial conditions $x3(\tau_3)=x2(\tau_3)$ from $t=\tau_3$ to $t=\tau_4$}
     \STATE{$P_M=\Pi2_{s3,s2}\Pi1_{s2,s1}\Pi0_{s1,s0}P0_{s0}$}
     \STATE{$\vec{x} = \vec{x3}(\tau_4)$}
     \STATE{$H_i(BIN(\vec{x}))=H_i(BIN(\vec{x}))+P_M H_{i-1}(j)$}
     \COMMENT{$BIN(\vec{x}) :$ bin containing $\vec{x}$}
     \ENDFOR
     \ENDFOR
     \ENDFOR
     \ENDFOR
     \ENDFOR
     \ENDFOR
  \end{algorithmic}
\end{algorithm}

In order to achieve this we consider a deterministic partition $\tau_0=0 <\tau_1 < \ldots < \tau_M=\tau$ of the interval $[0,\tau]$ such that $\Delta_M = \max_{j\in[1,M]} (\tau_{j} - \tau_{j-1})$ is small.
The main approximation of this method is to assume that $s_t$, for $t \in [0,\tau]$, is piecewise constant on this partition, more precisely, that $s_t = s_j := s_{\tau_j}, \, \text{for } t \in [\tau_j,\tau_{j+1}), \, 0 \leq j \leq M-1$.
This is rigorously true for intervals $[\tau_j,\tau_{j+1})$ completely contained between two successive random jump times of $s_t$. 
This situation becomes very frequent for a very fine partition (large $M$). 
Thus, the error generated by the approximation vanishes in the limit $M \to \infty$ (the rigorous result is Theorem~\ref{theorem1} given in the Results section).

\begin{figure}[!ht]
\centerline{\includegraphics[width=0.7\textwidth]{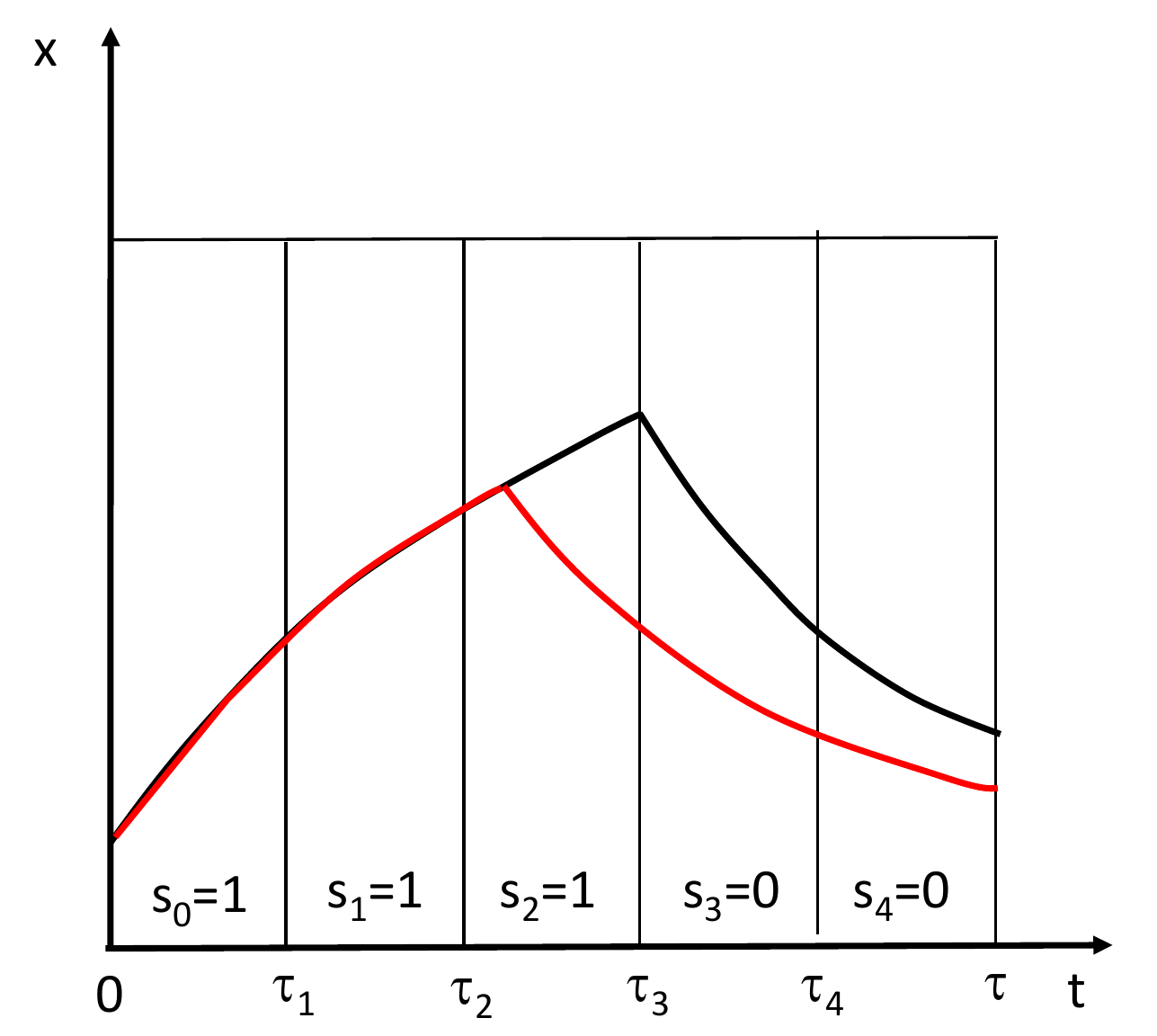}}
\caption{ \label{figure1}
Deterministic partition of time in the push forward method. 
Trajectories of the PDMP model (red line) having a jump of the discrete state inside $[\tau_2,\tau_3)$ are replaced by a trajectory (black line) having a jump of the discrete state at $\tau_3$. 
}
\end{figure}

For each path realization $S_M:=(s_0,s_1,\ldots,s_{M-1})\in \Omega :={\cal S}^M$ of the discrete states, we can compute $\vec{x}(t),\, t \in [0,\tau]$ as the continuous solution of the following ODE with piecewise-defined r.h.s:
\begin{equation} \label{eqxt}
\D{\vec{x}}{t} = \vec{V}_{s_j} (\vec{x}), \text{for }   t \in [\tau_j,\tau_{j+1}),\, 0\leq j \leq M-1 
\end{equation}
and with the initial condition $\vec{x}(0)=\vec{x}_0$.

In order to compute the probability $\proba{S_M}$ of a path realization we can use the fact that, given $\vec{x}_t$, $s_t$ is a finite state Markov process. Therefore,
\begin{equation}\label{SM}
\proba{S_M} = \Pi_{s_{M-1},s_{M-2}}(\tau_{M-2},\tau_{M-1}) \ldots \Pi_{s_1,s_0}(\tau_0,\tau_1) P_0^S(s_0),
\end{equation}
where $P_0^S: {\cal S} \to [0,1]$ is the initial distribution of the discrete state, and $\vec{\Pi}(\tau_{j},\tau_{j+1})$ is the solution, at $t=\tau_{j+1}$, of 
\begin{equation}\label{PI}
\D{\vec{\Pi}(\tau_{j},t)}{t} = H(\vec{x}_t) \vec{\Pi}(\tau_{j},t)
\end{equation}
with $\vec{\Pi}(\tau_{j},\tau_{j}) = {\mathbb I}$ and $\vec{x}_t$ is given by \eqref{eqxt}.

In order to compute the probability distribution of $\vec{x}$ at time $\tau$ one has to sum the contributions of all solutions of \eqref{eqxt}, obtained for the $N_s^{M}$ realisations of promoter state paths with weights given by the probabilities of the paths.

Suppose that we want to estimate the distribution of $\vec{x}(\tau)$, using a multivariate histogram with bin centers $\vec{c}^{(l_1,\ldots,l_N)}=(c_1^{l_1}, \ldots, c_N^{l_N}), \, 1\leq l_i \leq n_x,  1\leq i \leq N$ where $n_x$ is the number of bins in each direction $x_i,\, 1\leq i \leq N$. 
The initial distribution of $\vec{x}$ at time $t=0$ is given by the bin probabilities $p_0^{(l_1,\ldots,l_N)},1 \leq i \leq N, 1\leq l_i \leq  n_x$ and the distribution at time $\tau$ are given by the probabilities $p_\tau^{(l_1,\ldots,l_N)},1 \leq i \leq N, 1\leq l_i \leq  n_x$.

Let $\vec{x}^{(l_1,\ldots,l_N)}(t)$ be the solution of \eqref{eqxt} with $\vec{x}^{(l_1,\ldots,l_N)}(0)=\vec{c}^{(l_1,\ldots,l_N)}$ and let $(l_1',\ldots,l_N')$ be the histogram bin containing $\vec{x}^{(l_1,\ldots,l_N)}(\tau)$. The application $(l_1,\ldots,l_N) \to (l_1',\ldots,l_N'):= \psi(l_1,\ldots,l_N)$ is in general many to one. 
Given the probability $\proba{S_M}$ of a path $S_M \in \Omega$, the push forward distribution of $\vec{x}(\tau)$ is computed as  
\begin{equation}
p^{(l_1',\ldots,l_N')}_\tau = \sum_{S_M \in \Omega} \; \sum_{\psi(l_1,\ldots,l_N)=(l_1',\ldots,l_N')} p_0^{(l_1,\ldots,l_N)} \proba{S_M}, \label{distribution}
\end{equation}

The push-forward method can be applied recursively to compute the distribution at times $\tau, \, 2\tau, \ldots, n_t \tau$.
The complete algorithm is schematized in Algorithm~\ref{alg:pf} for the choice $M=4$. 

\begin{remark} \rm
The Algorithm~\ref{alg:pf} can be used for computing the full multivariate probability distribution of the vector $\vec{x}(t)$, but also for computing marginal distributions. 
For gene network the full multivariate distribution implies products from all genes, whereas a marginal distribution can select only one, or a small number of genes.
For marginals, the dimension $N$ is replaced by $N_m < N$ where $N_m$ is the number of components of interest of the vector $\vec{x}(t)$.
\end{remark}

\begin{remark} \rm
For gene networks, the numerical integration of the ODEs (steps 7,9,11 in the Algorithm~\ref{alg:pf}) can be replaced by symbolic solutions. 
For each path realization $S_M:=(\vec{s}_0,\vec{s}_1,\ldots,\vec{s}_{M-1})\in \Omega :=\{0,1,\ldots,2^N-1\}^M$ of promoter states, we can compute the protein and mRNA levels, $\vec{y}_t$ and $\vec{r}_t$, respectively, of all genes $i \in \{1,N\}$, at $t=\tau$:
\begin{eqnarray}
r^i_{\tau} &=& r^i_{0}e^{-\rho\tau} + \frac{k_0}{\rho} (1-e^{- \rho\tau}) + \frac{k_1-k_0}{\rho}\sum_{j=1}^{M-1}e^{-\rho\tau} (e^{-\rho\tau_{j+1}}-e^{-\rho\tau_j})s_j^i \label{eqrt} \\[2mm]
y^i_{\tau} &=& y^i_{0} e^{-a\tau} + \frac{b r^i_{0}}{a-\rho} (e^{-\rho\tau} - e^{-a\tau}) + \frac{bk_0}{\rho}\left(\frac{1-e^{-a\tau}}{a}+\frac{e^{-a\tau}-e^{-\rho\tau}}{a-\rho}\right) \notag \\
&+& \frac{b (k_1-k_0)}{\rho}e^{-a\tau}\sum_{j=1}^M s_{j-1}^i w_{j}, \label{eqyt}
\end{eqnarray}
for $i \in \{1,\ldots,N\}$.
Here,
\[
\begin{split}
w_{j} 
= & \frac{e^{(a-\rho)\tau}-e^{(a-\rho)\tau_j}}{a-\rho}(e^{\rho \tau_j}-e^{\rho \tau_{j-1}}) \\[2mm]
& - \frac{e^{(a-\rho)\tau_j}-e^{(a-\rho)\tau_{j-1}}}{a-\rho} \, e^{\rho \tau_{j-1}}+\frac{e^{a \tau_j}-e^{a \tau_{j-1}}}{a}
\end{split}
\]
with $s_j^i: = 0$ if promoter $i$ is OFF for $t \in [\tau_j,\tau_{j+1})$ and $s_j^i: = 1$ if promoter $i$ is ON for $t \in [\tau_j,\tau_{j+1})$.
\end{remark}

\subsubsection{Complexity issues}

The complexity of the push forward algorithm scales as $n_t n_x^N  N_s^M  (N+N_s^2)$, because there are $N_s^M$ possible paths $S_M$, $n_x^N$ histogram bins, and $n_t$ time complexity for solving \eqref{eqxt}, \eqref{PI} is $O (N+N_s^2)$. 
The complexity of computing marginal distributions of $N_m<N$ variables is lower and scales as $n_t n_x^{N_m}  N_s^M  (N+N_s^2)$.

\subsubsection{Mean field push-forward method}

A way to mitigate the computational burden of the Algorithm~\ref{alg:pf} is to use the mean field approximation. 
In the mean field approximation, the probabilities $\proba{S_M}$ are computed from averaged equations that are identical for each histogram bin.

More precisely, equation~\eqref{PI} is replaced by
\begin{equation}\label{PIMF}
\D{\vec{\Pi}(t',t)}{t} = \esp{H(\vec{x}_t)} \vec{\Pi}(t',t)
\end{equation}
where $t' \leq t$, $\vec{\Pi}(t',t') = {\mathbb I}$.

Using a Taylor expansion of $H(\vec{x}_t)$ around the expectation $\esp{\vec{x}_t}$, one gets
\begin{equation}\label{PPI}
\D{\vec{\Pi}(t',t)}{t} = 
H(\esp{\vec{x}_t}) + \frac{1}{2}\big( H''(\esp{\vec{x}_t}): 
\var(\vec{x}_t)\big) \vec{\Pi}(t',t),
\end{equation}
where $\var(\vec{x}_t)$ is the variance/covariance matrix of $\vec{x}_t$, $H''$ is the element-wise second derivative matrix of $H$ and $:$ stands for the double dot product. 

In \eqref{PPI} the moments $\esp{\vec{x}_t}$ and $\var(\vec{x}_t)$ are either available analytically or are solutions of ODEs obtained using moment closures such as in \cite{singh2010stochastic}. 

The main advantage of the mean-field approximation is that the transition matrix elements can be computed outside the discrete variables loop.
The complete algorithm is presented in the Algorithm~\ref{alg:pfmf}. 
The time complexity in this case is $n_t n_x^N  N_s^M  N$, because in the innermost loop one has to solve only \eqref{eqxt}, whose time complexity is $O (N)$.


For gene networks, the mean field push-forward procedure can be applied also to ODE dynamics of the individual genes. 
This is possible because during ON or OFF periods, the ODE dynamics of one gene is uncoupled from that of another gene. 
Furthermore, the $N_s\times N_s$ transition matrix $\vec{H}$ can be replaced by $N/2$ $2\times 2$ transition matrices of one gene with elements averaged over the values of the other genes.  
This approximation reduces the complexity of the calculations to $2 n_t (n_x)^2 N 2^M$, which is linear in the number of genes.
The mean field push-forward algorithm for gene networks is presented in the Algorithm~\ref{alg:pfmfgn}.

\enlargethispage{5mm}

This method has already been used for particular models. 
In \cite{innocentini2018time} we have replaced the regulation term $f_2 y_1(t)$ occurring in the transition matrix of the gene network model $M_1$
by its mean $f_2 \esp{y_1(t)}$. 
This means that the gene $2$ switches between its ON and OFF states with rates given by the mean of the regulatory protein $y_1$. 
In this case both $\vec{H}$ and $\vec{\Pi}$ can be computed analytically, which leads to a drastic reduction in the execution time. 
This simple mean field approach is suitable for the model $M_1$, which contains only linear regulation terms.
For non-linear regulation terms, $\vec{\Pi}$ can not generally be computed analytically. 
Moreover, the naive mean field approach introduces biases.
For instance, in the case of the model $M_2$, the approximation $f_2 y_1(t)/(K_1+y_1(t)) \approx f_2 \esp{y_1(t)}/(K_1+\esp{y_1(t)})$ is poor. In the CMSB2019 paper we proposed a better approximation \cite{innocentini2019effective}, in which we replace $f_2 y_1(t)/(K_1+y_1(t))$ by its mean value and use
\begin{equation}
\esp{\frac{f_2 y_1(t)}{K_1+y_1(t)}} \approx  \frac{f_2 \esp{y_1(t)}}{(K_1+\esp{y_1(t)})} - \frac{f_2 }{(K_1+\esp{y_1(t)})^3} \operatorname{Var}(y_1(t)),
\end{equation}
in order to correct the bias. This approach is generalized by \eqref{PPI}.


\newpage

\vfill

\begin{algorithm}[!ht]
  \begin{algorithmic}[1]
    \caption{\label{alg:pfmf}$\operatorname{PDMPpushforwardmeanfield}$}
    \REQUIRE 1.~ $p_0^g(x,s)$, initial distribution of $(x_0,s_0)$. 2.~$\tau$ time step for computing the time-dependent distribution 
\smallskip
    \ENSURE  $n_t$ vectors $H_0,\ldots,H_{nt}$ representing the distribution of $x$ at times $0,\tau,\ldots,n_t \tau$
    \smallskip
    \STATE{compute $H_0,P0$ from initial distribution}
     \FOR{$i:=1$ \TO $n_t$}
          \FOR{$s0:=1$ \TO $N_s$}
               \FOR{$s1:=1$ \TO $N_s$}
                    \FOR{$s2:=1$ \TO $N_s$}
                        \FOR{$s3:=1$ \TO $N_s$}
                 \STATE{Compute \par $P_M(s_0,s_1,s_2,s_3)=\Pi_{s3,s2}(\tau_2,\tau_3)\Pi_{s2,s1}(\tau_1,\tau_2)\Pi_{s1,s0}(\tau_0,\tau_1)P0_{s0}$
                 where $\Pi(t',t)$ is the solution of \eqref{PPI}.
                 }
                        \ENDFOR
                     \ENDFOR
                \ENDFOR
           \ENDFOR
     \FOR{$s0:=0$ \TO $1$}
     \FOR{$j:=1$ \TO $n_x^N$}
          \STATE{solve $\D{\vec{x0}}{t} = V_{s_0}(\vec{x0})$,  with initial conditions $\vec{x0}(0)=C_j$, from $t=0$ to $t=\tau_1$}
     \FOR{$s1:=0$ \TO $1$}
          \STATE{solve $\D{\vec{x1}}{t} = V_{s_1}(\vec{x1}) $ with initial conditions $x1(\tau_1)=x0(\tau_1)$ from $t=\tau_1$ to $t=\tau_2$}
     \FOR{$s2:=0$ \TO $1$}
          \STATE{solve $\D{\vec{x2}}{t} = V_{s_2}(\vec{x2})$ with initial conditions $x2(\tau_2)=x1(\tau_2)$ from $t=\tau_2$ to $t=\tau_3$}
     \FOR{$s3:=0$ \TO $1$}
     \STATE{solve $\D{\vec{x3}}{t} = V_{s_3}(\vec{x3})$ with initial conditions $x3(\tau_3)=x2(\tau_3)$ from $t=\tau_3$ to $t=\tau_4$}
     \STATE{$\vec{x} = \vec{x3}(\tau_4)$}
     \STATE{$H_i(BIN(\vec{x}))=H_i(BIN(\vec{x}))+P_M(s_0,s_1,s_2,s_3) H_{i-1}(j)$
     where $BIN(\vec{x})$ is the bin containing $\vec{x}$}
     \ENDFOR
     \ENDFOR
     \ENDFOR
     \ENDFOR
     \ENDFOR
     \ENDFOR
  \end{algorithmic}
\end{algorithm}

\vfill

\newpage

\vfill

\begin{algorithm}[!ht]
  \begin{algorithmic}[1]
    \caption{\label{alg:pfmfgn}$
    \operatorname{PDMPpushforwardmeanfieldgenenetwork}$}
    \REQUIRE 1.~for each gene $g\in \{1,2,\ldots,N/2\}$, $p_0(x,s)$, initial distribution of $(x_0,s_0)$, where $x=(r,p)$. 2.~$\tau$ time step for computing the time-dependent distribution 
\smallskip
    \ENSURE for each gene $g \in \{1,2,\ldots,N/2\}$, $n_t$ vectors $H_0,\ldots,H_{nt}$ representing the distribution at times $0,\tau,\ldots,n_t \tau$
    \smallskip
    \FOR{$g:=1$ \TO $N/2$}
    \STATE{compute $H_0,P0$ from initial distribution for gene $g$}
     \FOR{$i:=1$ \TO $n_t$}
          \FOR{$s0:=0$ \TO $1$}
               \FOR{$s1:=0$ \TO $1$}
                    \FOR{$s2:=0$ \TO $1$}
                        \FOR{$s3:=0$ \TO $1$}
                 \STATE{Compute \WRP $P_M(s_0,s_1,s_2,s_3)=\Pi_{s3,s2}(\tau_2,\tau_3)\Pi_{s2,s1}(\tau_1,\tau_2)\Pi_{s1,s0}(\tau_0,\tau_1)P0_{s0}$
                 where $\Pi(t',t)$ is the solution of \eqref{PPI} for gene $g$.
                 }
                        \ENDFOR
                     \ENDFOR
                \ENDFOR
           \ENDFOR
     \FOR{$s0:=1$ \TO $2$}
     \FOR{$j:=1$ \TO $n_x^2$}
          \STATE{solve $\D{\vec{x0}}{t} = V_{s_0}(\vec{x0})$,  with initial conditions $\vec{x0}(0)=C_j$, from $t=0$ to $t=\tau_1$, and $V_s$ is defined by \eqref{standard1}  for gene $g$}
     \FOR{$s1:=1$ \TO $N_s$}
          \STATE{solve $\D{\vec{x1}}{t} = V_{s_1}(\vec{x1}) $ with initial conditions $x1(\tau_1)=x0(\tau_1)$ from $t=\tau_1$ to $t=\tau_2$}
     \FOR{$s2:=1$ \TO $N_s$}
          \STATE{solve $\D{\vec{x2}}{t} = V_{s_2}(\vec{x2})$ with initial conditions $x2(\tau_2)=x1(\tau_2)$ from $t=\tau_2$ to $t=\tau_3$}
     \FOR{$s3:=1$ \TO $N_s$}
     \STATE{solve $\D{\vec{x3}}{t} = V_{s_3}(\vec{x3})$ with initial conditions $x3(\tau_3)=x2(\tau_3)$ from $t=\tau_3$ to $t=\tau_4$}
     \STATE{$\vec{x} = \vec{x3}(\tau_4)$}
     \STATE{$H_i(BIN(\vec{x}))=H_i(BIN(\vec{x}))+P_M(s_0,s_1,s_2,s_3) H_{i-1}(j)$
     where $BIN(\vec{x})$ is the bin containing $\vec{x}$}
     \ENDFOR
     \ENDFOR
     \ENDFOR
     \ENDFOR
     \ENDFOR
     \ENDFOR
     \ENDFOR
  \end{algorithmic}
\end{algorithm}

\vfill

\newpage

As in \cite{innocentini2018time} we can use analytic expressions for $\esp{y_1(t)}$, but also for $\var(y_1(t))$. 
These expressions can be found in the \ref{appA}. 
Although the elements of matrix $\vec{H}$ have analytic expressions, the elements of the matrix $\vec{\Pi}$ contain integrals that must be computed numerically. For the model $M_2$, we have
\begin{equation} \setlength{\arraycolsep}{2.5pt} 
\vec{\Pi}_1(\tau,\tau') =
\begin{bmatrix}
(1-p_{1,on}) + p_{1,on}e^{-\epsilon_1 (\tau'-\tau)} & 
(1-p_{1,on}) (1 - e^{-\epsilon_1 (\tau'-\tau)}) \\[1mm]
p_{1,on} (1 - e^{-\epsilon_1 (\tau'-\tau)}) & 
p_{1,on} + (1-p_{1,on})e^{-\epsilon_1 (\tau'-\tau)}  \\
\end{bmatrix},
\end{equation}
for the transition rates of the first gene, where $p_{1,on} = f_1/(f_1+h_1)$, $\epsilon_1 = (f_1+h_1)/\rho$, and
\begin{equation}
\begin{aligned}&
\vec{\Pi}_2(\tau,\tau') = 
\begin{bmatrix}
K(\tau,\tau') + h_2 \int_{\tau}^{\tau'} K(t,\tau') \, dt &
h_2 \int_{\tau}^{\tau'} K(t,\tau') \, dt \\[1mm]
1 - K(\tau,\tau') - h_2 \int_{\tau}^{\tau'} K(t,\tau') \, dt &
1 - h_2 \int_{\tau}^{\tau'} K(t,\tau') \, dt
\end{bmatrix},
\end{aligned}\label{pi2}
\end{equation}
for the transitions of the second gene, where $K(\tau,\tau')=e^{-\int_{\tau}^{\tau'}(h_2 + F_2(t))\, dt}$ and $F_2(t) = f_2 \esp{\frac{y_1(t)}{K_1+y_1(t)}}$.


\section{Results}

\subsection{Convergence of the push-forward method}\label{sec:convergence}
The probability distribution obtained with the push-forward method converges to the exact PDMP distribution in the limit $M \to \infty$. This is a consequence of the following theorem:

\begin{theorem}\label{theorem1}
Let $\Phi_{S_M}(t,\vec{x})$ be the flow defined by the formulas \eqref{eqxt}, such that $\vec{x}_t=\Phi_{S_M}(t,\vec{x}(0))$ for $t\in[0,\tau]$. 
Let $\mu_t^M : \mathcal{B}(\R^{N})\to \R_+$ be the probability measure defined as $\mu_t^M (A) = \sum_{S_M \in  \Omega} \proba{S_M} \mu_0(\Phi_{S_M}^{-1}(t,A))$, where $\mu_0 : \mathcal{B}(\R^{N})\to \R_+$ is the probability distribution of $\vec{x}$ at $t=0$, $\proba{S_M}$ is given by \eqref{SM}, and $\mathcal{B}(\R^{N})$ is the $\sigma$-algebra of Borel sets on $\R^{N}$.
Let $\mu_t$, the exact distribution of $\vec{x}_t$ for the PDMP defined in Section \ref{s2.2}, with initial values $(\vec{x}_0,s_0)$ distributed according to $\mu_0 \times P_0^S$.
Assume that the vector fields $\vec{V}_s$, $s\in S$, and the transition matrix $\vec{H}$ are $C^1$ functions and that there exists a bounded set of $\R^N$ such that all flows $\Phi_s$, $s\in S$, leave $B$ invariant. 
Assume that $|\tau_{i}-\tau_{i-1}| < C/M$ for all $i\in [1,M]$, where $C$ is a positive constant.
Then, for all $t\in [0,\tau]$, $\mu_t^M$  converges in distribution to  $\mu_t$, when $M \to \infty$. 
More precisely, for all Lipschitz functions $\varphi$ on $\R^N$, there  exists a constant $\kappa$ depending on the data of the PDMP, $\tau$ and the Lipschitz constant of $\varphi$ such that:
\[
\left|\int_{B}\varphi(x) \mu_t^M(dx) -\int_{B}\varphi(x) \mu_t(dx)\right|= |\mathbb{E}(\varphi(\Phi_{S_M}(t,\vec{x}))) - \mathbb{E}(\varphi(\vec{x}_t))|\le \kappa/M .
\]
Also for all  Borel set $A$ in $\mathcal{B}(\R^{N})$ such that $\mu_t(\partial A)=0$ we have $\mu^M_t(A)\to \mu_t(A)$ when $M\to \infty$. 
Here, $\partial A$ denotes the boundary of $A$.
\end{theorem}

The proof of this theorem is given in \ref{appB}. 
It is inspired by the classical proof of weak order of convergence for the Euler scheme for a stochastic differential equation (see \cite{talay}). 

\begin{remark} \rm
If the flows $\Phi_{s}$ are not known explicitly, a numerical scheme can be used. This introduces another source of error. Our proof easily extends and provide a similar result of convergence.
If one wants to investigate the convergence of $p^{(l_1,\ldots,l_N)}_\tau$ given by \eqref{distribution},  this follows from the above theorem only under the assumption that the probability that the PDMP is on the boundary of the bins at time $t$ is $0$. Note this is not restrictive and happens only in pathological situations.
Also, if the initial distribution has a smooth density, we can prove that the error estimate above holds for borelian bounded functions $\varphi$, thus we can choose $\varphi$ to be an indicator function and obtain error bounds for these probabilities without this restriction. 
More precisely, we have for any Borel set $A$:
\[
\left|\mu_t^M(A)-\mu_t(A)\right|\le \kappa /M,
\]
where now $\kappa$ depends on the Lipschitz constant of the initial density, see Remark \ref{r} in \ref{appB}.
\end{remark}

\subsection{Testing the performance of various methods}

\subsubsection{Testing accuracy and speed of
push-forward method compared
to the Monte-Carlo method}
In order to test the push-forward method, we compared the resulting probability distributions with the ones obtained by the Monte-Carlo method using the direct simulation of the PDMP.
We considered the models $M_1$ and $M_2$ with the following parameters: $\rho=1$, $p_1=\frac{f}{f+h}=1/2$, $a=1/5$, $b=4$, $k_0=4$, $k_1=40$ for the two genes.
For the parameter $\epsilon=\frac{f+h}{\rho}$ we used two values: $\epsilon=0.5$ for slow genes and $\epsilon=5.5$ for fast genes. 
We tested the slow-slow and the fast-fast combinations of parameters.

The initial distribution of the promoters state was $P_0^S( (0,0) )=1$ where the state $(0,0)$ means that both promoters are OFF.
The initial probability measure $\mu_0$ was a delta Dirac distribution centered at $r^1 = r^2=0$ and $y^1=y^2=0$. 
This initial condition is obtained by always setting the direct simulation of the PDMP to start from $r_1(0)=r_2(0)=0$, $y_1(0)=y_2(0)=0$, and $s^1_{0}= s^2_{0} = 0$.
The simulations were performed between $t_0=0$ and $t_{max}=20$ for fast genes and between $t_0=0$ and $t_{max}=90$ for slow genes. In order to estimate the distributions we have used $MC=50000$ samples for the highest sampling.
The push-forward method was implemented with $M=10$ equal length sub-intervals of $[0,\tau]$.
The time step $\tau$ was chosen to be $\tau=2$ for fast genes and $\tau=15$ for slow genes. 
The procedure was iterated $10$ times for fast genes (up to $t_{max}=20$) and $6$ times for slow genes (up to $t_{max}=90$).

\begin{figure}[!ht]
\includegraphics[width=0.5\textwidth]{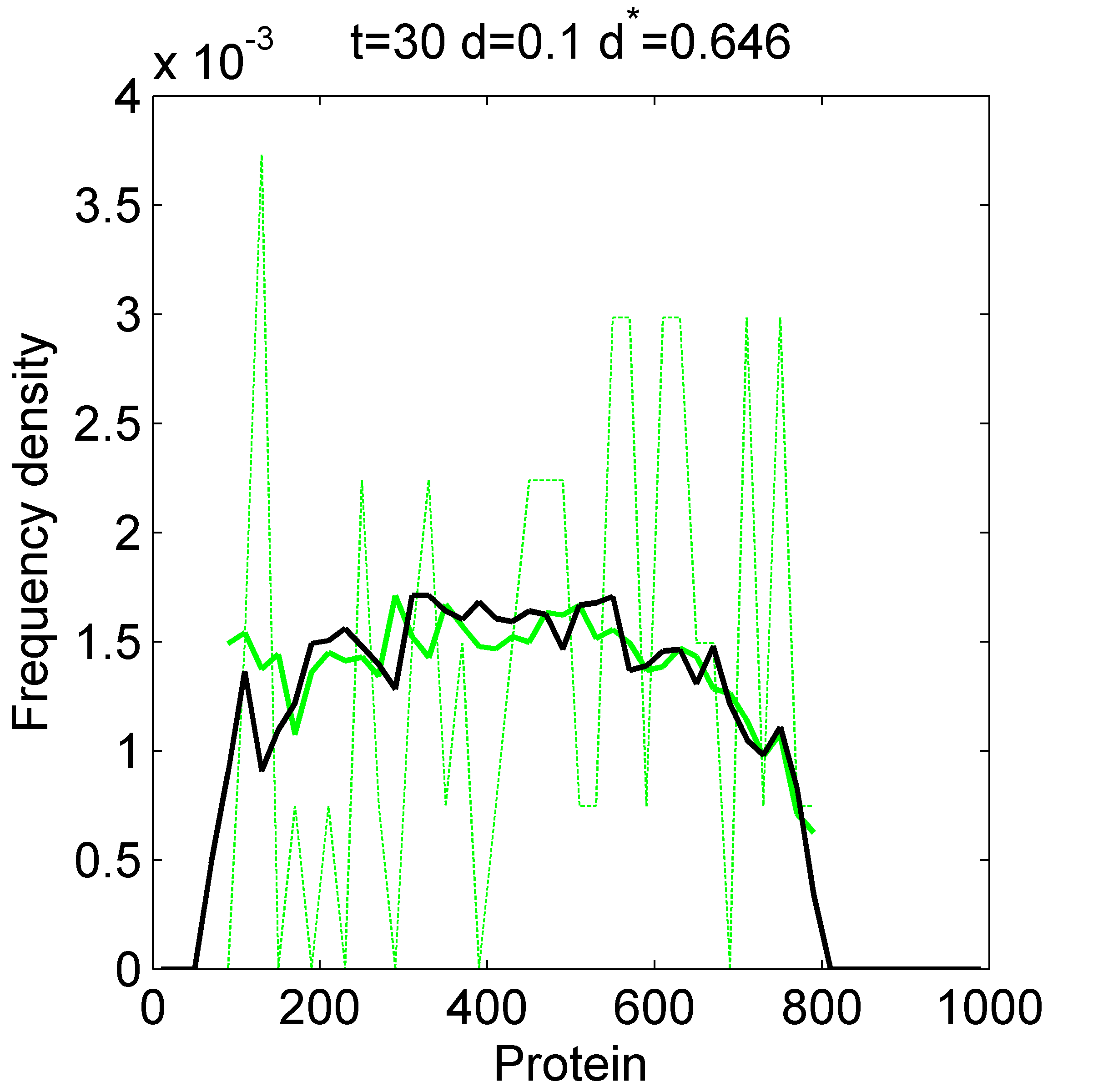}
\includegraphics[width=0.5\textwidth]{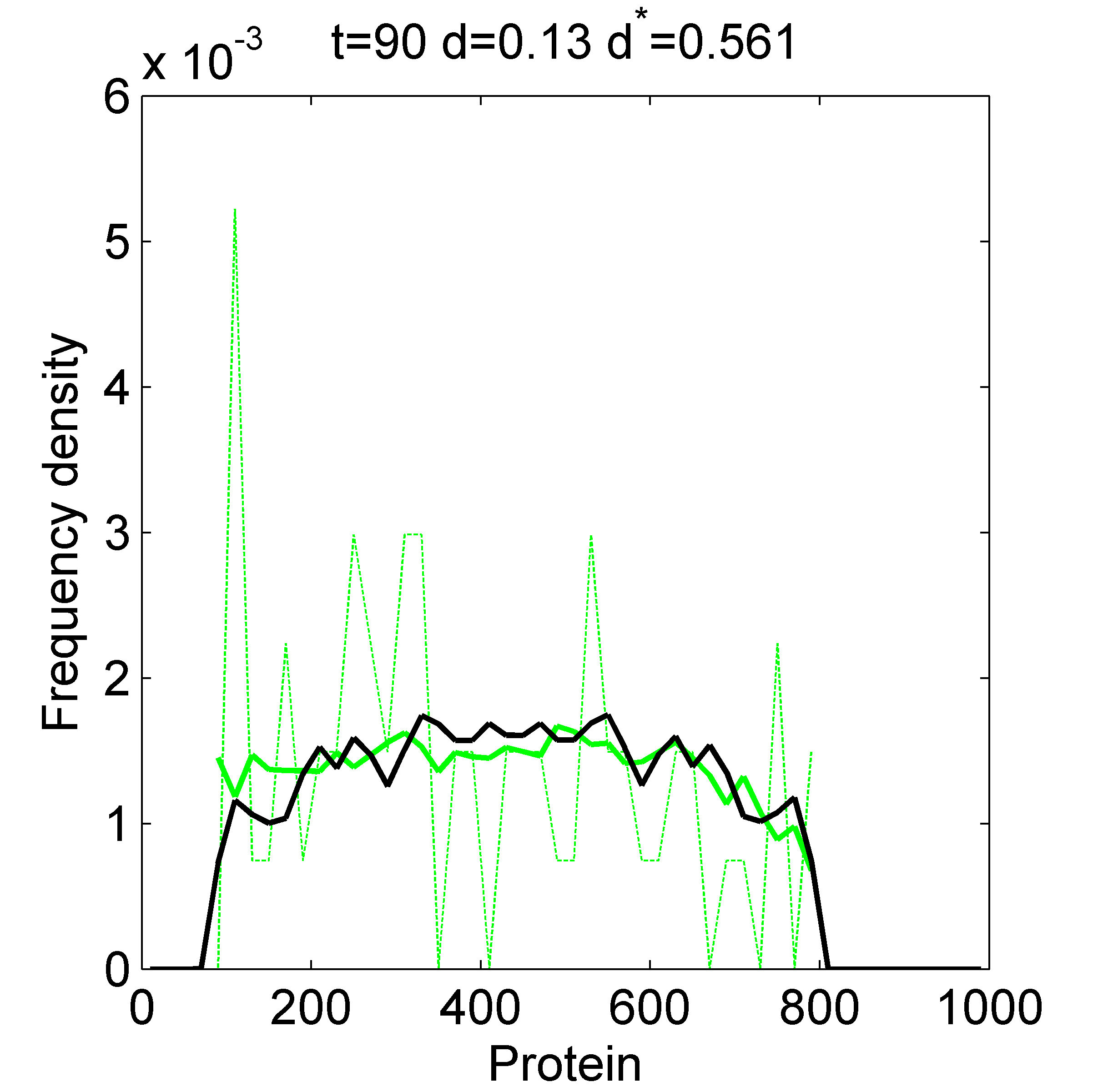}
\includegraphics[width=0.5\textwidth]{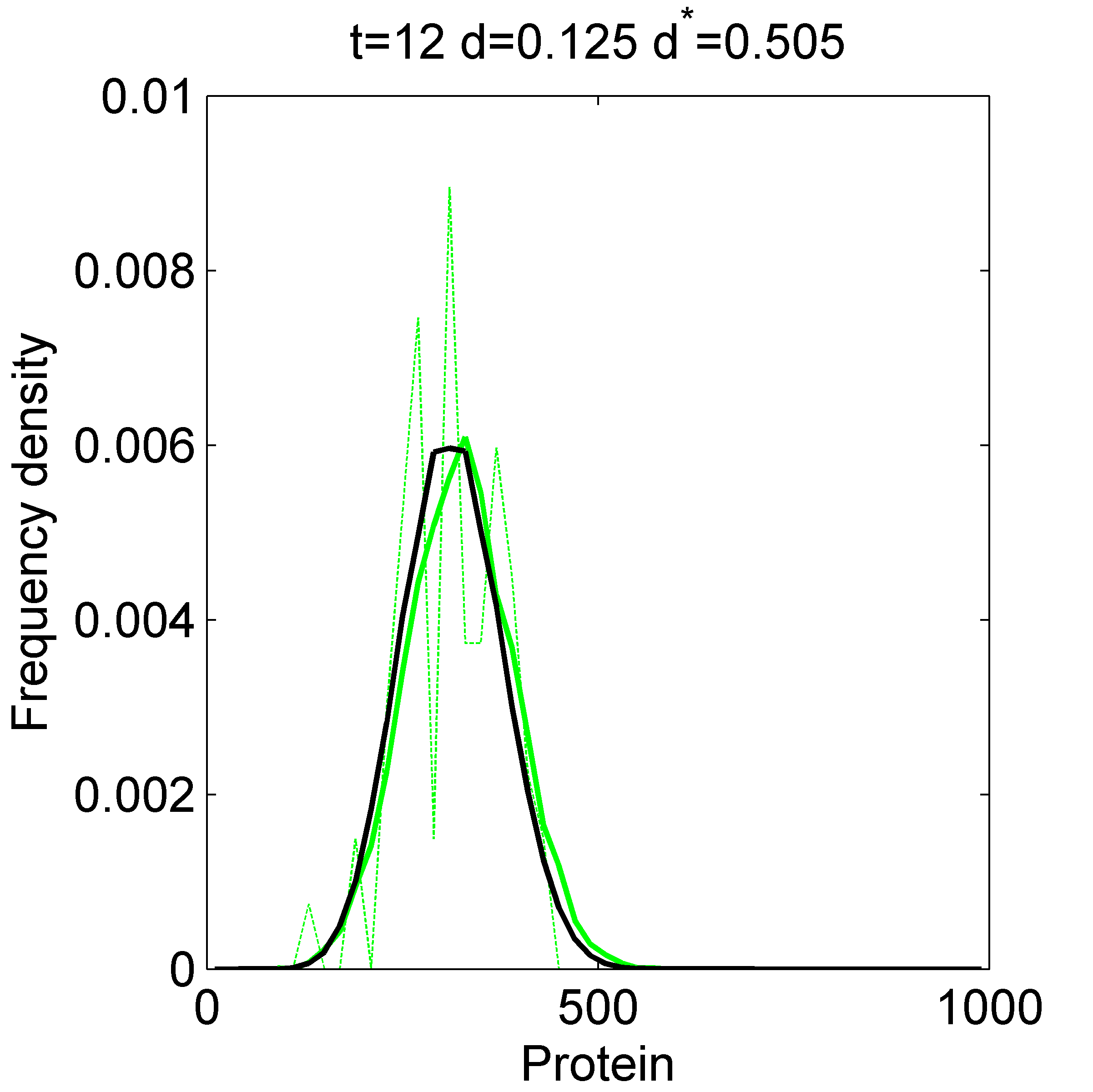}
\includegraphics[width=0.5\textwidth]{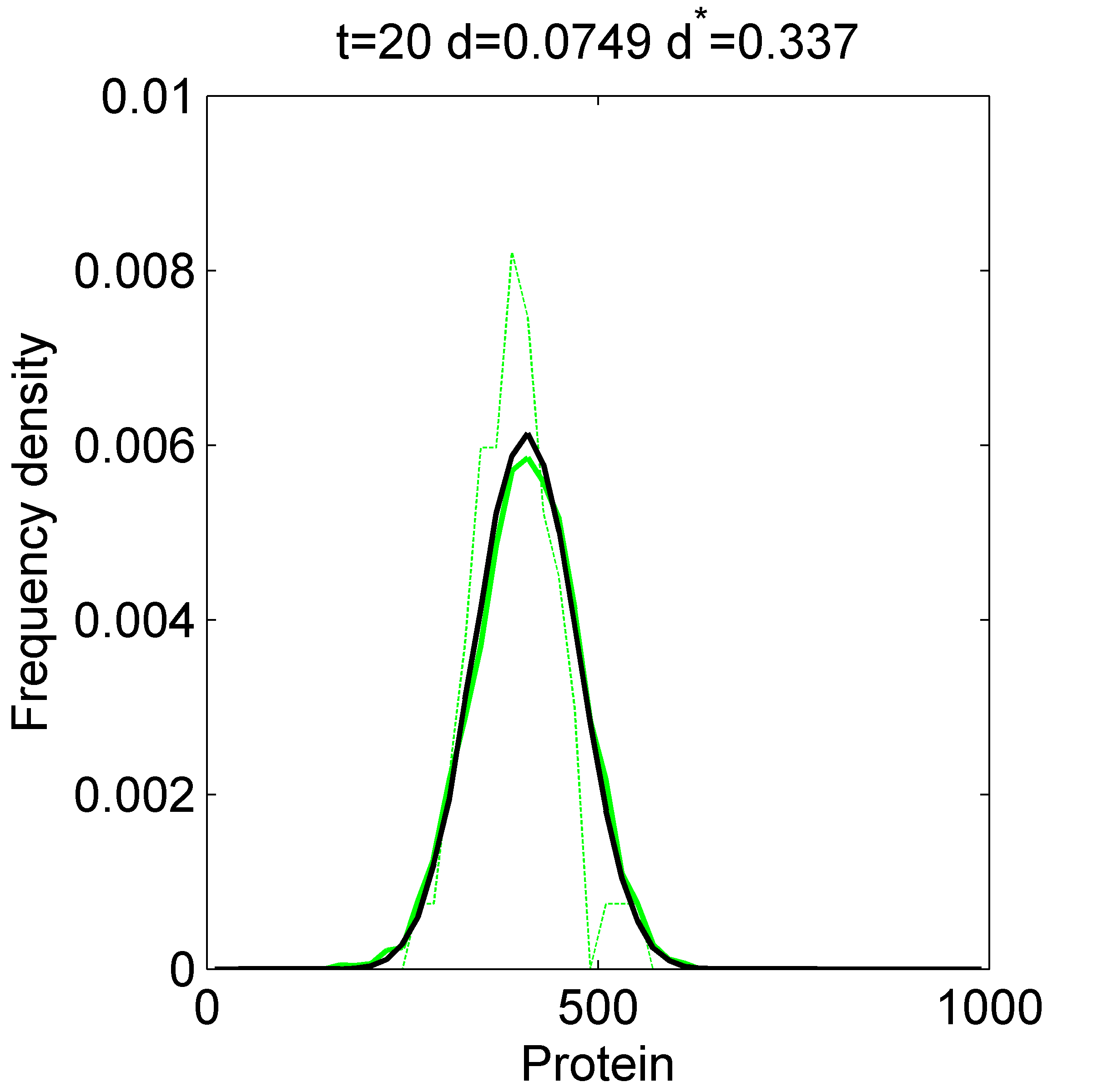}

\vspace{-4mm}
\caption{ \label{fig1}
Histograms of protein  for the second gene, produced by the Monte-Carlo method (green lines) and by the Push-forward method (black lines) for the model $M_1$.
The green dotted line results from low sampling Monte-Carlo with similar execution time as the push-forward method, whereas the solid green line results from high sampling Monte-Carlo.
The distances, defined by \eqref{distance}, are between low sampling and high sampling Monte-Carlo ($d^*$) and between push-forward and high sampling Monte-Carlo ($d$).
}
\vspace{-3mm}
\end{figure}

\begin{figure}[!ht]
\includegraphics[width=0.5\textwidth]{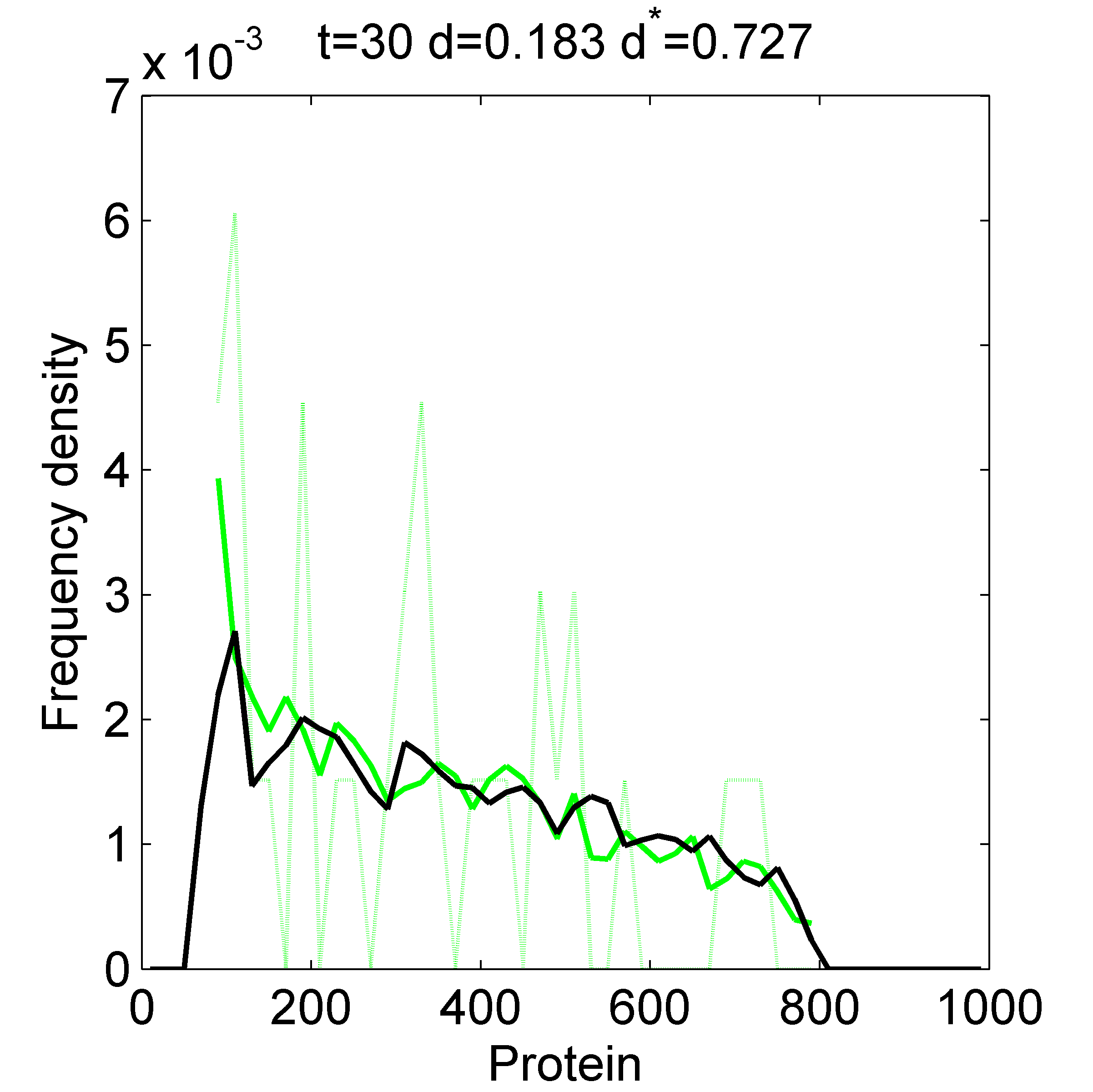}
\includegraphics[width=0.5\textwidth]{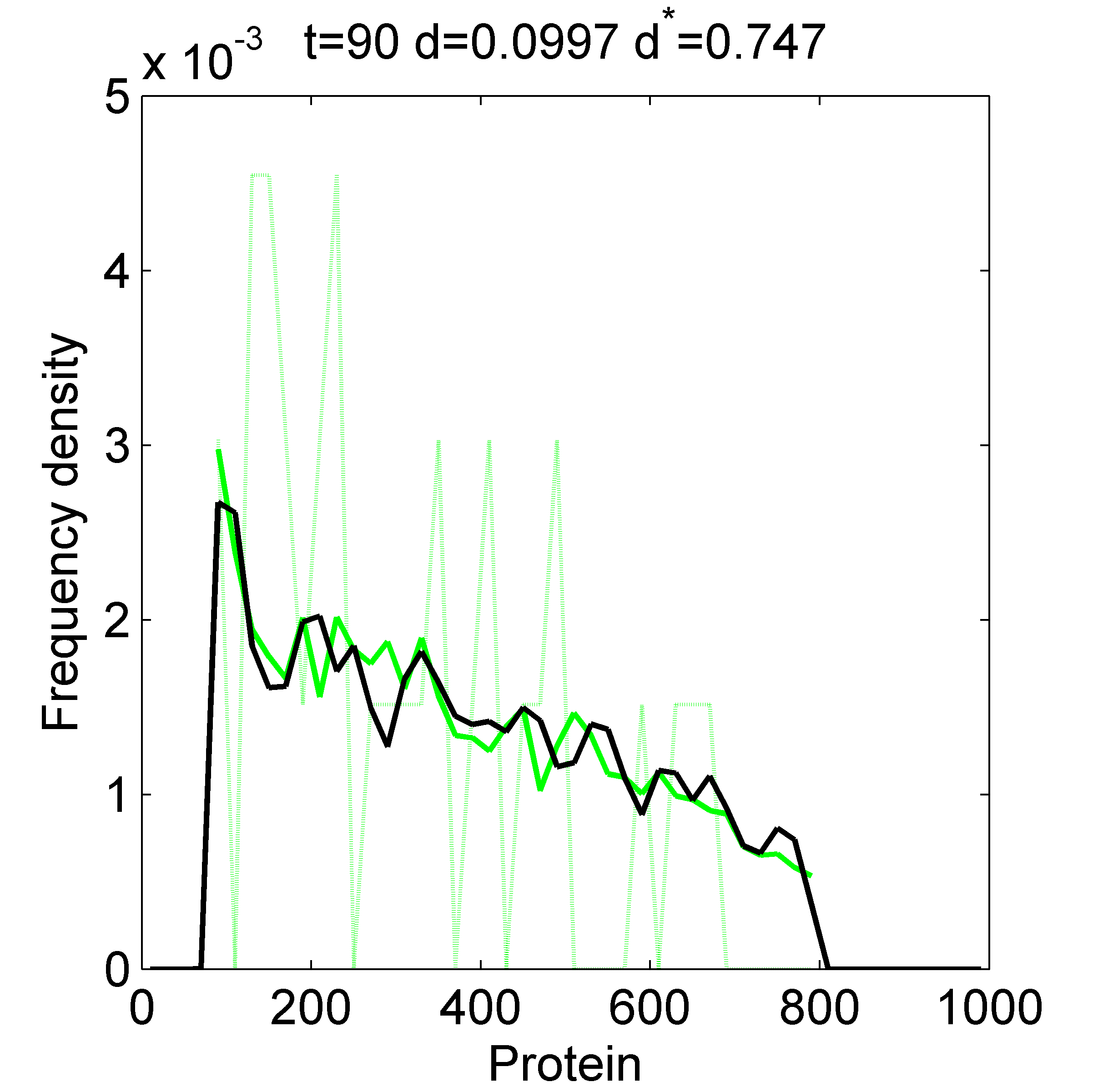}
\includegraphics[width=0.5\textwidth]{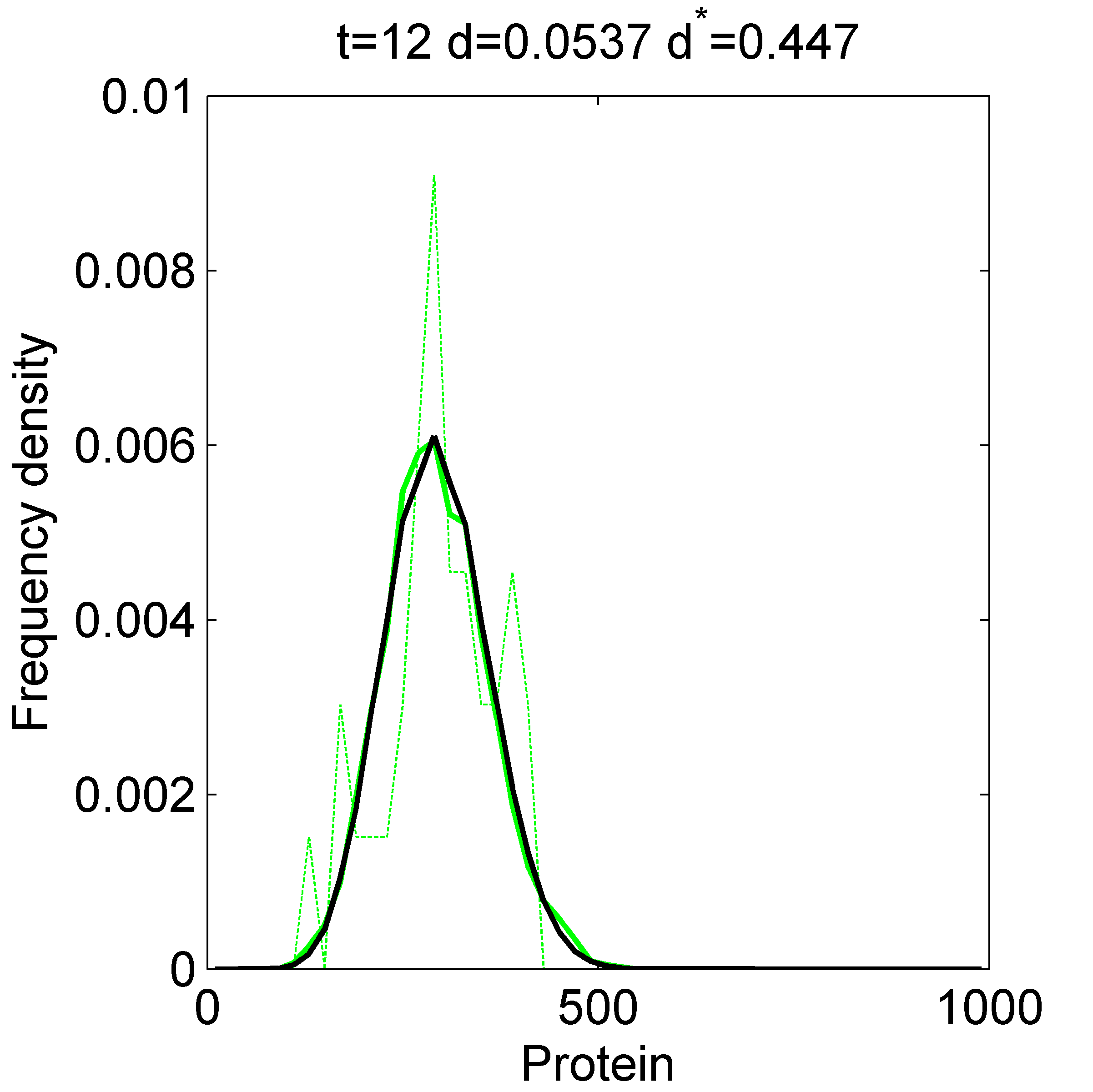}
\includegraphics[width=0.5\textwidth]{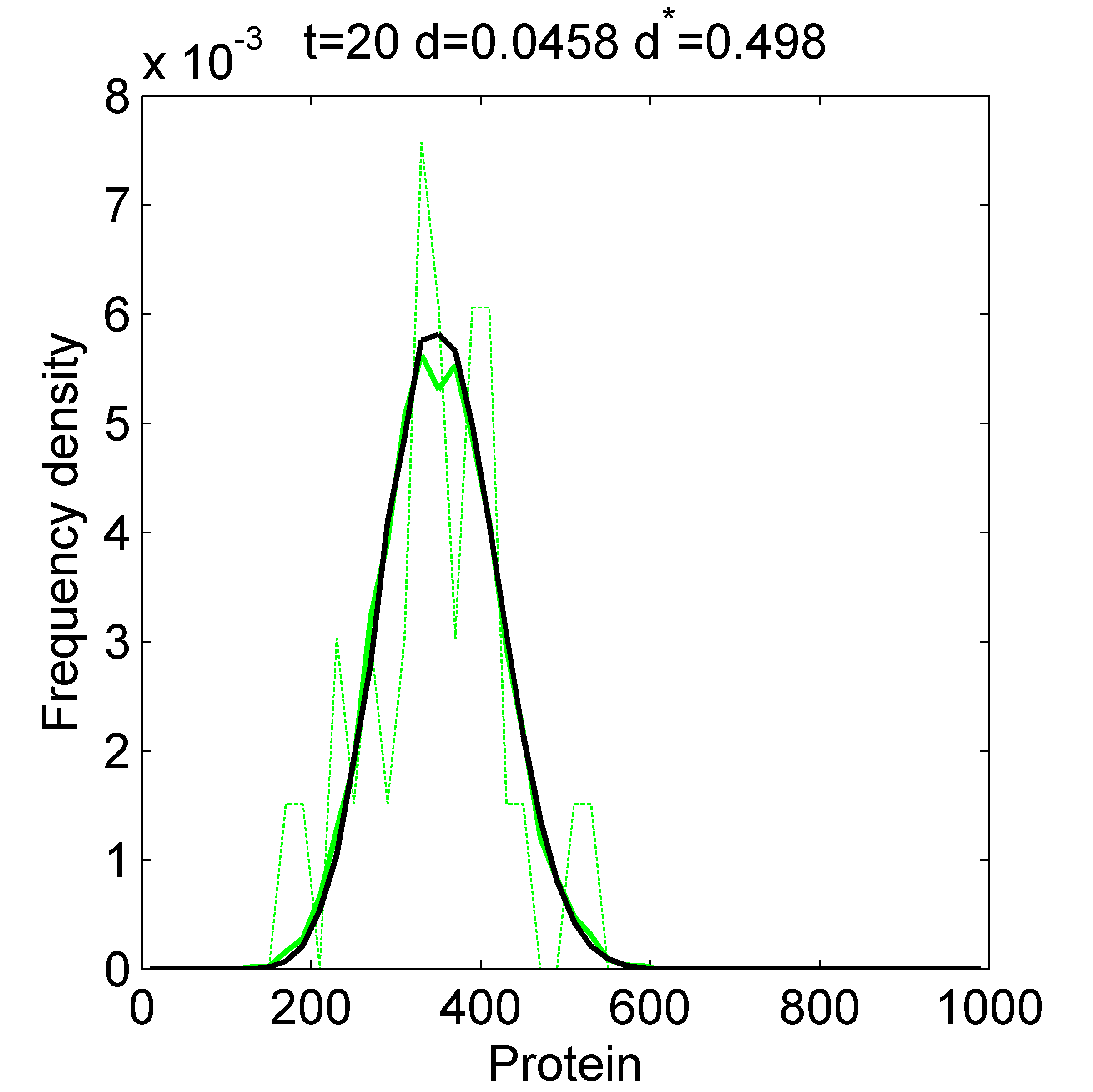}
\caption{ \label{fig2}
Histograms of protein  for the second gene, produced by the Monte-Carlo method (green lines) and by the Push-forward method (black lines) for the model $M_2$.
The green dotted line results from low sampling Monte-Carlo with similar execution time as the push-forward method, whereas the solid green line results from high sampling Monte-Carlo.
The distances, defined by \eqref{distance}, are between low sampling and high sampling Monte-Carlo ($d^*$) and between push-forward and high sampling Monte-Carlo ($d$).
}
\end{figure}

The execution times are shown in the Table~\ref{table1}.
The comparison of the probability distributions are illustrated in the Figures~\ref{fig1} and \ref{fig2}.
In order to quantify the relative difference between methods we use the $L^1$ distance between distributions. More precisely, if $p(x)$ and $\tilde{p}(x)$ are probability density functions to be compared, the distance between them is
\begin{equation} \label{distance}
d = \int | p(x) - \tilde{p}(x) | \,dx.
\end{equation}

\begin{table}[!ht]
\begin{center}
\begin{tabular}{|c|c|c|}
\hline
Model  &  Monte-Carlo high sampling [min]  & Push-forward [s]  \\
\hline
$M_1$ slow-slow & 45     &  20 \\
$M_1$  fast-fast & 74    &  30 \\
$M_2$  slow-slow & 447   &  20 \\
$M_2$  fast-fast & 758   &  30 \\
\hline
\end{tabular}
\end{center}
\vspace{-4mm}
\caption{\label{table1} 
Execution times for different methods.
All the methods were implemented in Matlab R2013b running on a single core (multi-threading inactivated) of a Intel i5-700u 2.5 GHz processor.
The Monte-Carlo method computed the next jump waiting time
(Algoritm~\ref{alg:ns}) using the analytical solution of \eqref{T1} for $M_1$ and the numerical solution of  \eqref{T1} for $M_2$. 
The push-forward method used 
Algorithm~\ref{alg:pfmfgn} and
analytic solutions for mRNA and protein trajectories from \eqref{eqyt}, \eqref{eqxt} and numerical computation of the integrals in \eqref{pi2}, for both models.
}
\vspace{-1mm}
\end{table}

This distance was computed for distributions resulting from the push-forward method and the Monte-Carlo method with the highest sampling. We have also used a reduced sampling Monte-Carlo scheme whose execution time is similar to the one of the push-forward method. The distributions resulting from low sampling and high sampling Monte-Carlo were compared using the same distance.
Figures~\ref{fig1} and \ref{fig2} clearly show that, for the same execution time, the push-forward method outperforms the Monte-Carlo method.

\pagebreak

\subsubsection{Comparing the Monte Carlo, push-forward and Liouville-master methods}

Using Python, we have implemented Algorithm~\ref{alg:pf} for the one gene model, to compute mRNA $(r(t))$ and protein $(y(t))$ probability densities $P(r,t)$ and $P(y,t)$, respectively. For the one gene model, we have considered two switching regimes: slow $(\epsilon=0.5)$ and fast $(\epsilon=5.5)$. Also in Python, we have implemented Algorithm~\ref{alg:pfmf} for the two genes model, to compute the probability densities for mRNA and protein associated to gene one ($P(r_1,t)$ and $P(y_1,t)$) and for mRNA and protein associated to gene two ($P(r_2,t)$ and $P(y_2,t)$). For the two genes model, we have considered four different switch configurations: slow-slow, fast-fast, slow-fast and fast-slow. For all implementations each time interval $[0,\tau]$ has been partitioned into four sub-intervals of equal sizes $(M=4)$ resulting in the sequence $\{s_{0}, s_{1}, s_{2}, s_{3}\}$, representing the state of the switch inside each sub-interval ($s=s_{j}:=s_{t_{j}}$ for $t\in[t_j,t_{j+1}), j=0,...,M-1$), leading to $2^{4}$ path realizations. For all slow genes we have set $t_0=0$, $t_4=9$ and evaluated the solution up to $t_{max}=90$, using the composition rule between successive time intervals. For all fast genes we have set $t_0=0$, $t_4=1$ and evaluated the solution up to $t_{max}=20$, using the composition rule between successive time intervals, as well.

\begin{table}[!ht]
\begin{center}
\begin{tabular}{|c|c|c|c|c|}
\hline
Model  &  Monte-Carlo  & PF & LME \\
       & (high sampling)&   &  \\
\hline
One gene slow       &  -   & 2.55  & 97.39 \\
One gene fast       &  -   & 5.12  & 21.64 \\
Two genes slow-slow &  45  & 6.86  & 97.39 \\
Two genes fast-fast &  74  & 10.85 & 21.64 \\
Two genes slow-fast & 243  & 9.22  & 21.64 \\
Two genes fast-slow & 249  & 8.83  & 97.39 \\
\hline
\end{tabular}
\end{center}
\caption{\label{table1_time} 
Execution times (in minutes for all cases) for different methods to compute the probability distributions for one gene model and two genes model.
Methods: PF = Push-forward, LME = Liouville-master equation}
\end{table}

\begin{figure}[!ht]
\includegraphics[width=0.46\textwidth]{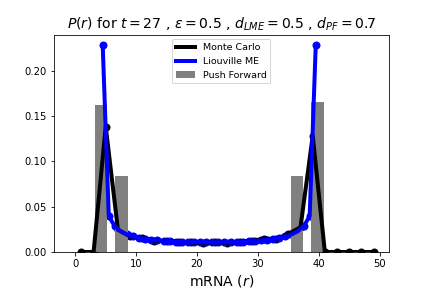}
\includegraphics[width=0.46\textwidth]{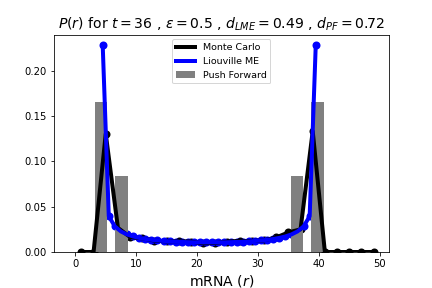}
\includegraphics[width=0.46\textwidth]{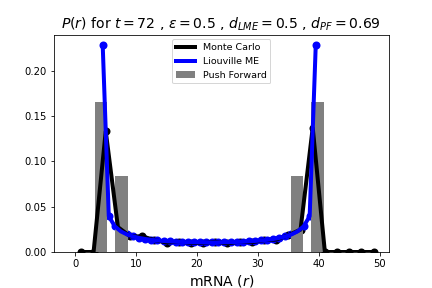}
\includegraphics[width=0.46\textwidth]{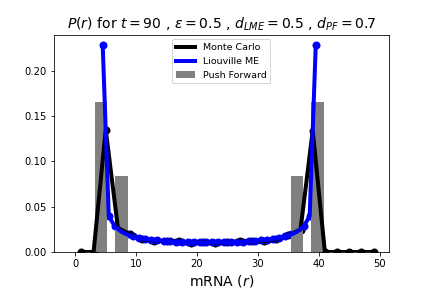}
\includegraphics[width=0.46\textwidth]{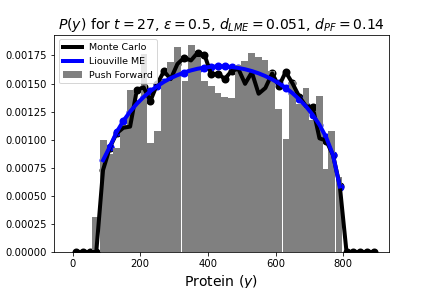}
\includegraphics[width=0.46\textwidth]{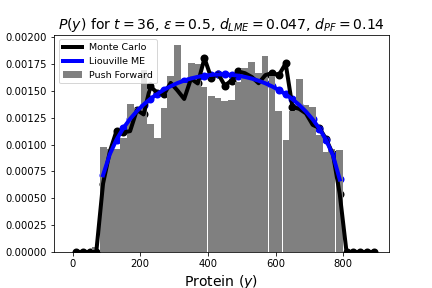}
\includegraphics[width=0.46\textwidth]{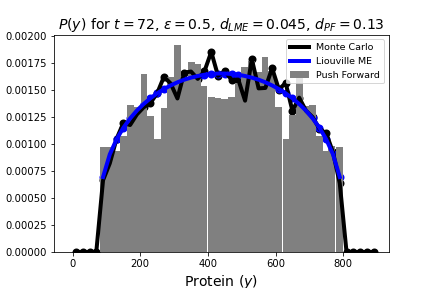}
\hfill
\includegraphics[width=0.46\textwidth]{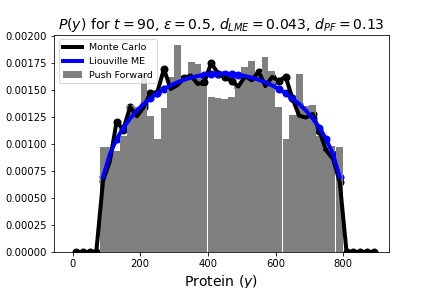}
\caption{\label{gfig1} 
This set of plots shows the comparison between the computed probability distribution for mRNA ($P(r,t)$) and protein ($P(y,t)$), using the push forward method (Algorithm~\ref{alg:pf}), direct Monte-Carlo simulation (Algorithms~\ref{alg:ns} and \ref{alg:mc}) and numerical solution of the Liouville-master Equation (Algorithm~\ref{alg:pde}). The model is for one gene in the slow switch regime ($\epsilon=0.5$).
The execution times of PF is $2.55$ minutes in Python and $97.39$ minutes for LME in MATLAB. The values $d_{LME}$ and $d_{PF}$ are shown in each plot and measure the distances between the LME equation and MC simulation, PF and MC simulation, respectively, as given in \eqref{distance}. The value of parameters are: $p_0=0.5$, $k_0=4$, $k_1=40$, $a=1/5$ and $b=4$. For initial conditions we set: $r(0)=y(0)=0$ and $p_{0}(0)=1$.}
\end{figure}

\begin{figure}[!ht]
\includegraphics[width=0.46\textwidth]{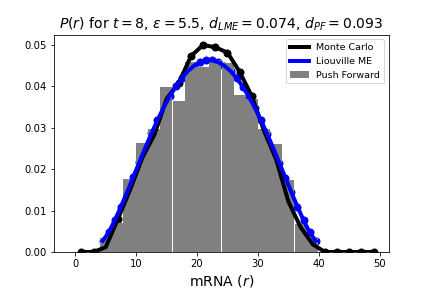}
\includegraphics[width=0.46\textwidth]{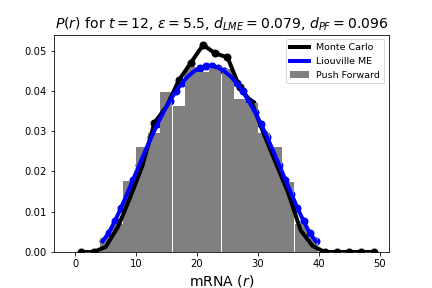}
\includegraphics[width=0.46\textwidth]{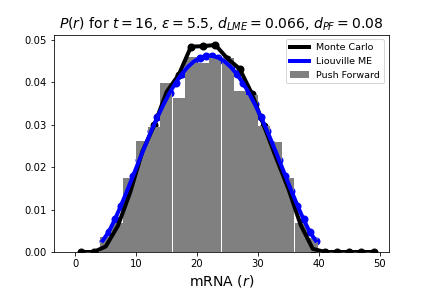}
\includegraphics[width=0.46\textwidth]{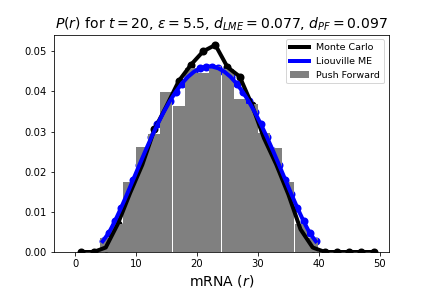}
\includegraphics[width=0.46\textwidth]{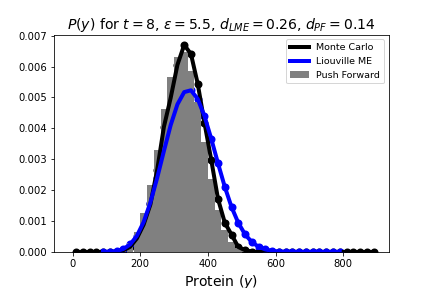}
\includegraphics[width=0.46\textwidth]{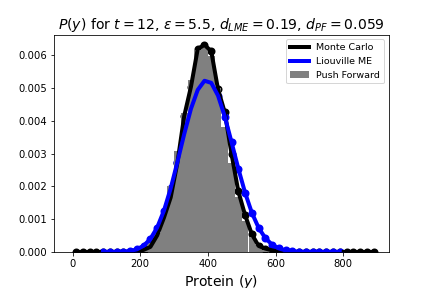}
\includegraphics[width=0.46\textwidth]{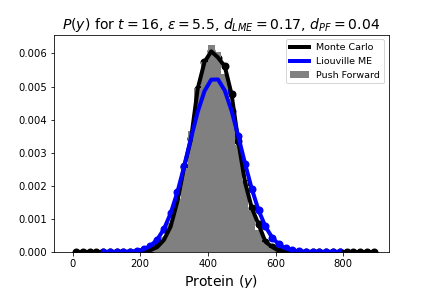}
\hfill
\includegraphics[width=0.46\textwidth]{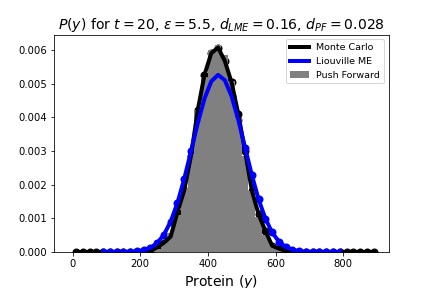}
\caption{\label{gfig2} 
This set of plots shows the comparison between the computed probability distribution for mRNA ($P(r,t)$) and protein ($P(y,t)$), using PF (Algorithm~\ref{alg:pf}), MC simulation (Algorithms~\ref{alg:ns} and \ref{alg:mc}) and numerical solution LME (Algorithm~\ref{alg:pde}). The model is for one gene producing in the fast switch regime ($\epsilon=5.5$). The execution time of the PF is $5.12$ minutes in Pyhton, and $21.64$ minutes for LME in MATLAB. The values $d_{LME}$ and $d_{PF}$ are shown in each plot and measure the distances between the LME equation and MC simulation, and PF and MC simulation, respectively, as given in \eqref{distance}. The value of parameters are: $p_0=0.5$, $k_0=4$, $k_1=40$, $a=1/5$ and $b=4$. For initial conditions we set: $r(0)=y(0)=0$ and $p_{0}(0)=1$.}
\end{figure}

\begin{figure}[!ht]
\includegraphics[width=0.46\textwidth]{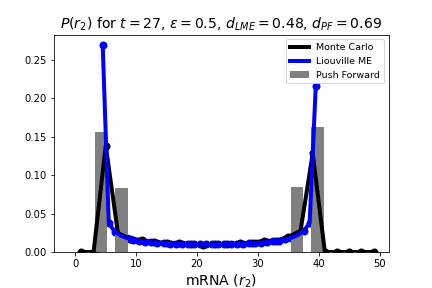}
\includegraphics[width=0.46\textwidth]{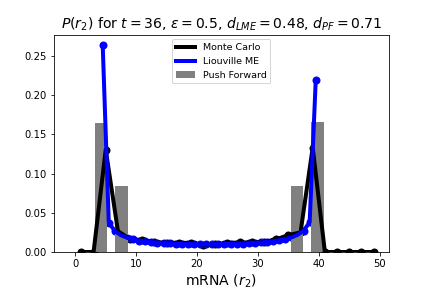}
\includegraphics[width=0.46\textwidth]{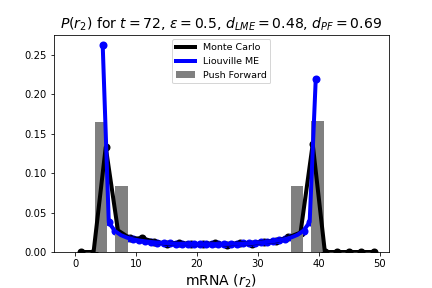}
\includegraphics[width=0.46\textwidth]{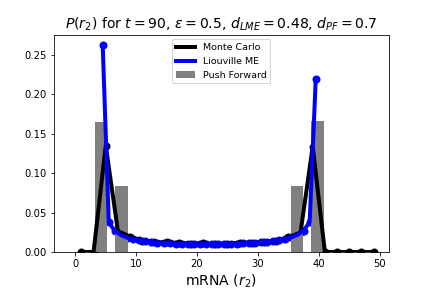}
\includegraphics[width=0.46\textwidth]{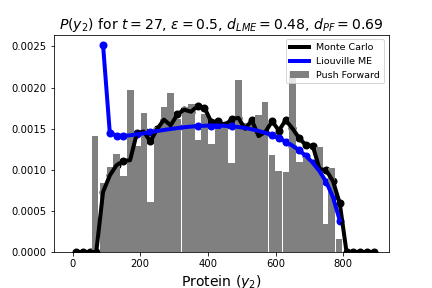}
\includegraphics[width=0.46\textwidth]{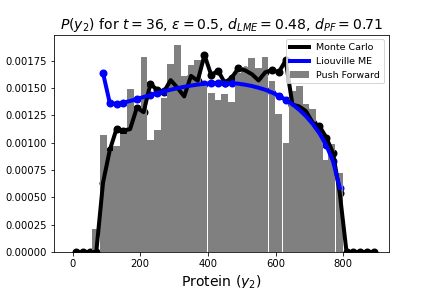}
\includegraphics[width=0.46\textwidth]{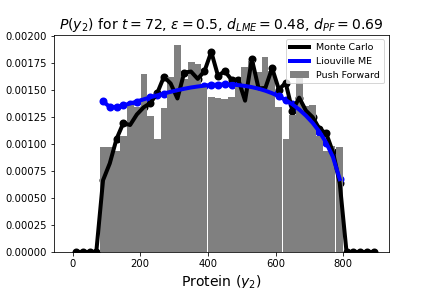}
\hfill
\includegraphics[width=0.46\textwidth]{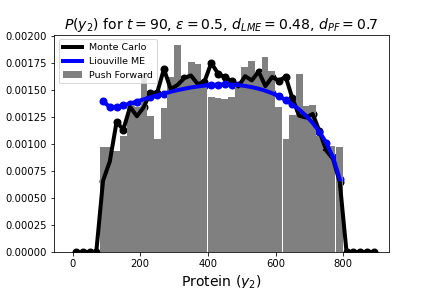}
\caption{\label{gfig3} 
This set of plots shows the comparison between the computed probability density for mRNA ($P(r_2,t)$) and protein ($P(y_2,t)$) associated to gene two, using PF (Algorithm~\ref{alg:pfmf}), MC simulation (Algorithms~\ref{alg:ns} and \ref{alg:mc}) and numerical solution LME (Algorithm~\ref{alg:pde}). The model is $M_1$ in the slow-slow regime ($\epsilon=0.5$). The execution time of PF is $6.86$ minutes in Python, and $97.39$ minutes for LME in MATLAB. The values $d_{LME}$ and $d_{PF}$ are shown in each plot and measure the distances between the LME equation and MC simulation, and PF and MC simulation, respectively, as given in \eqref{distance}. The value of parameters, for both genes are: $p_0=0.5$, $k_0=4$, $k_1=40$, $a=1/5$ and $b=4$. For initial conditions we set: $r_1(0)=y_1(0)=r_2(0)=y_2(0)=0$ and $p_{0}(0)=1$.}
\end{figure} 

\begin{figure}[!ht]
\includegraphics[width=0.46\textwidth]{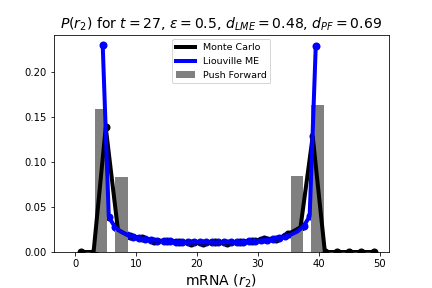}
\includegraphics[width=0.46\textwidth]{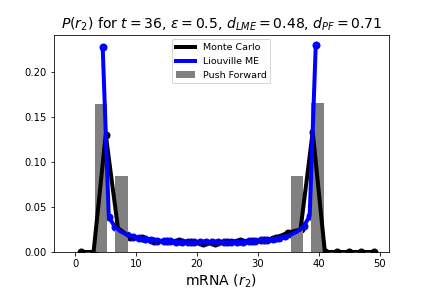}
\includegraphics[width=0.46\textwidth]{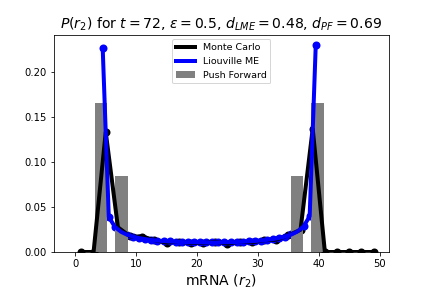}
\includegraphics[width=0.46\textwidth]{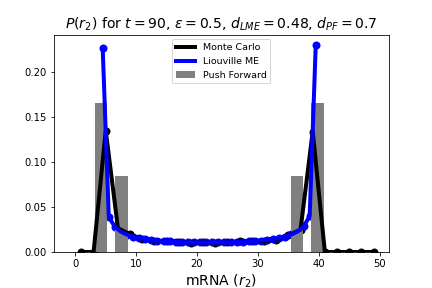}
\includegraphics[width=0.46\textwidth]{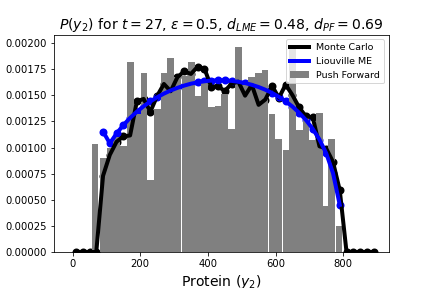}
\includegraphics[width=0.46\textwidth]{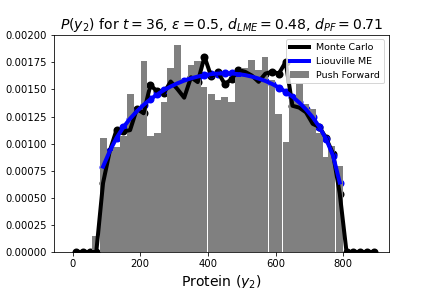}
\includegraphics[width=0.46\textwidth]{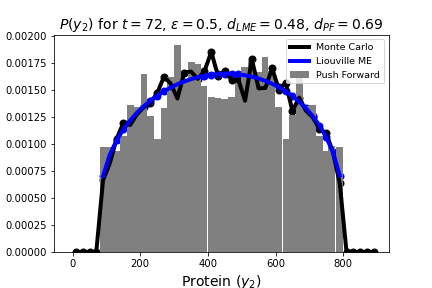}
\hfill
\includegraphics[width=0.46\textwidth]{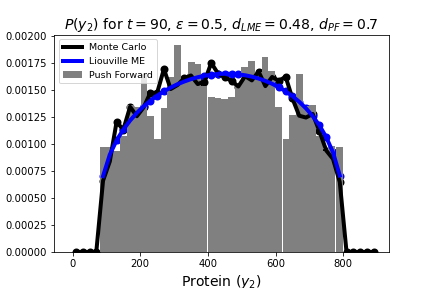}
\caption{\label{gfig4} 
This set of plots shows the comparison between the computed probability density for mRNA ($P(r_2,t)$) and protein ($P(y_2,t)$) associated to gene two, using PF (described in Algorithm~\ref{alg:pfmf}), MC simulation (Algorithms~\ref{alg:ns} and \ref{alg:mc}) and numerical solution LME (Algorithm~\ref{alg:pde}). The model is $M_1$ in the fast-slow regime ($\epsilon_1=5.5$ and $\epsilon_2=0.5$). The execution time of PF is $8.83$ minutes in Python and $97.39$ minutes for LME in MATLAB. The values $d_{LME}$ and $d_{PF}$ are shown in each plot and measure the distances between the LME equation and MC simulation, and PF and MC simulation, respectively, as given in \eqref{distance}. The value of parameters, for both genes are: $p_0=0.5$, $k_0=4$, $k_1=40$, $a=1/5$ and $b=4$. For initial conditions we set: $r_1(0)=y_1(0)=r_2(0)=y_2(0)=0$ and $p_{0}(0)=1$.}
\end{figure} 

We have compared the results obtained by the implementation of the push forward method (PF), as described above, with  Monte-Carlo simulation (MC) (Algorithms~\ref{alg:ns} and \ref{alg:mc}) and with numerical solution of the Liouville-master equation (LME) (Algorithm~\ref{alg:pde}). The comparison between execution times for each model and each method can be found in Table~\ref{table1_time} (all the execution times are expressed in minutes). 

In order to illustrate our results we have produced Figures~\ref{gfig1}, ~\ref{gfig2}, ~\ref{gfig3} and \ref{gfig4}, for selected models and for methods: Push forward (PF), Monte-Carlo simulation (MC) and Liouville-master equation (LME).

The one gene model (Figures~\ref{gfig1} and \ref{gfig2}) has as initial conditions $p_{0}(0)=1$, where $p_{0}(0)$ is the probability to find the switch in state $0$ at time $t=0$, and $r(0)=y(0)=0$, for both switch regimes: slow and fast.
The two genes model (Figures~\ref{gfig3} and ~\ref{gfig4}) has as initial conditions $p_{0}^{1}(0)=p_{0}^{2}(0)=1$ (meaning that at time $t=0$ both genes are in state $0$ with probability one), $r_{1}(0)=y_{1}(0)=r_{2}(0)=y_{2}(0)=0$, for both switch configurations: slow-slow and slow-fast. 
We used the same parameters for all the models and for all the genes: $\epsilon_{1}=\epsilon_{2}=0.5$ (slow switch), $\epsilon_{1}=\epsilon_{2}=5.5$ (fast switch),  $p_{0}^{1}=p_{0}^{2}=0.5$, $\rho_{1}=\rho_{2}=1$, $k_{1}(\sigma_1=0)=k_{2}(\sigma_2=0)=4$, $k_{1}(\sigma_1=1)=k_{2}(\sigma_2=1)=40$, $a_{1}=a_{2}=1/5$ and $b_{1}=b_{2}=4$. We have quantified the relative difference between the distributions obtained by two distinct methods using the $L^1$ distance \eqref{distance}.
The distance between the distributions obtained by the Push Forward method and the Monte-Carlo method (high sampling) is indicated in the plots by $d_{PF}$, and the distance between the distributions obtained by the Liouville-master equation and Monte-Carlo method (high sampling) is indicated by $d_{LME}$.

\section{Discussion and conclusion}

Combining direct simulation of PDMP gene network models and analytic formulae for the ODE flow provides an effective, easy to implement method for computing time dependent, multivariate probability distributions of these models. 
However, the precision of the Monte-Carlo estimates of the distributions increases with $\sqrt{MC}$, where $MC$ is the number of Monte-Carlo samples. 
For this reason, the execution time of the Monte-Carlo method, although smaller when compared to PDMP simulation methods, which implement numerical resolution of the ODEs, such as reported in \cite{lin2018efficient} (data not shown), is large when compared to the push-forward method. 
The push-forward method represents an effective alternative to Monte-Carlo and Liouville-master equation methods, ensuring reduced execution time. 

With respect to an earlier implementation of this method for gene networks in \cite{innocentini2016protein} we used promoter states instead of mRNA copy numbers as discrete variables of the PDMP. As a consequence, the number of discrete states is lower and we can afford to increase the number $M$ of temporal subdivisions. 
Compared to the similar work done in \cite{innocentini2018time} we used second moments of the protein distribution, which took into account the correlation of the promoter states and lead to increased accuracy in the case of nonlinear regulation.
Although the protein moments and the exponential transition rate matrix $\vec{\Pi}$ can be computed numerically, the effectiveness of the push-forward method is increased when analytic expressions are available for these quantities. 
In this paper, these expressions were computed for particular cases.

The push-forward method is an approximate method, and its accuracy relies on the careful choice of the temporal and spatial step densities, namely of the integers $M$, $n_t$, $n_x$. We have rigorously proved that the  convergence of the push-forward method is of order one, with errors that scale with $1/M$.

We situate our findings in the broader effort of the community to produce new effective simulation algorithms for computational biology. Although in this manuscript we restricted ourselves to examples of modeling gene networks, the push forward method
and the algorithms of this paper
can be applied to a broader class of phenomena where some continuous quantity is perturbed by a discrete stochastic process. 
A very interesting and important example is the stochastic representation of ion channel kinetics in neuron models \cite{anderson2015ionchannel}. The system of equations describing the kinetics of an ion channel is mathematically equivalent to the system of equations describing our one gene model. More complex neuron models with multiple types of ion channels (stochastic versions of Hodgkin-Huxley or Morris-Lecar deterministic models, for instance) are also covered by our PDMP formalism and algorithms. Thus, the push forward method as presented in Algorithm \ref{alg:pf} can be used to obtain the histograms for the membrane voltage for rather general neuron models. A detailed comparison between the push forward method and the one presented in \cite{anderson2015ionchannel} will be the subject of a future work.

\appendix

\section*{Appendices}

\section{Expectation and variance of the protein}
\label{appA}


For the sake of completeness, we give explicit formulas for the expectation and the variance of the protein synthesized by a constitutive promoter (gene 1 of models $M_1$ and $M_2$).
The details of the calculation can be found in \cite{innocentini2019effective}.

The expectation is given by
\begin{equation}
\esp{y_t} = M_0+M_1 e^{-at}+ M_2e^{-\rho t}+ M_3e^{-\epsilon t},
\label{a16}
\end{equation}
where
\begin{eqnarray*}
M_0 &=& \frac{b(k_0 + (k_1-k_0) p_1)}{a}, \\
M_1 &=& \esp{x_{0}} - \frac{b\esp{y_{0}}}{a-\rho}+ \frac{bk_0}{a(a-\rho)}
+ \frac{b(k_1-k_0)(p_{10}-p_1)}{(a-\rho\epsilon)(a-\rho)}+\frac{b(k_1-k_0)p_1}{a(a-\rho)}, \\
M_2 &=& \frac{b\esp{y_{0}}}{a-\rho} - \frac{bk_0}{a-\rho} - \frac{b(k_1-k_0)(p_{10}-p_1)}{\rho(1-\epsilon)(a-\rho)} - \frac{b(k_1-k_0)p_1}{a-\rho}, \\
M_3 &=& \frac{b(k_1-k_0)(p_{10}-p_1)}{\rho(1-\epsilon)(a-\epsilon \rho)}.
\label{a17}
\end{eqnarray*}

The variance is given by
\begin{equation}
\begin{split}
&\var(y_t) = 
\var(y_{0}) e^{-2at} + b^2 \var(r_{0}) \left( \frac{e^{-\rho t}-e^{-a t}}{a-\rho} \right)^2 \\
& - \left[\frac{(p_{10}-p_1)(k_1-k_0)b}{\rho(1-\epsilon)}\left( \frac{e^{-\rho\epsilon t}-e^{-a t}}{a-\rho\epsilon} - \frac{e^{-\epsilon t}-e^{-a t}}{a-\rho\epsilon}\right) \right]^2 \\
& + \frac{p_1(1-p_1)(k_1-k_0)^2 b^2}{\rho^2}
(V_0+V_1 e^{-(a+\rho\epsilon)t} + V_2e^{-\rho(1+\epsilon)t}+ V_3 e^{-2at} + V_4 e^{-(a+\rho)t}+ V_5 e^{-2\rho t}) \\
& + \frac{(1-2p_1)(p10-p_1)(k_1-k_0)^2 b^2}{\rho^2}
(V_6 e^{-\rho \epsilon t}+V_7 e^{-(a-\rho \epsilon) t}+V_8 e^{-\rho (\epsilon+1) t}+V_9 e^{-(a+\rho \epsilon) t}+V_{10} e^{-\rho t}+V_{11} e^{-2\rho t}),
\end{split}
\end{equation}
where
\begin{align*}
V_0 &= \frac{a+(\epsilon+1)\rho}{a(a+\rho\epsilon)(a+\rho)(\epsilon+1)}, &
V_1 &= -\frac{2}{(a^2-\rho^2\epsilon^2)(a-\rho)(\epsilon-1)}, \\
V_2 &= \frac{2}{(a-\rho\epsilon)(a-\rho)(\epsilon^2-1)}, &
V_3 &= \frac{1}{a(a-\rho\epsilon)(a-\rho)^2}, \\ 
V_4 &=\frac{2(a+(1-2\epsilon)\rho)}{(a-\rho\epsilon)(a-\rho)^2(a+\rho)(\epsilon-1)}, &
V_5 &= -\frac{1}{(\epsilon-1)(a-\rho)^2}, \\
V_6 &= -\frac{2(2a+(2-\epsilon)\rho)}{a(\epsilon-2)(2a-\rho\epsilon)(a+(1-\epsilon)\rho)}, &
V_7 &= \frac{2}{(1-\epsilon)a(a-\rho)(a-\rho\epsilon)}, \\
V_8 &= \frac{2}{(a-\rho\epsilon)(a-\rho)(\epsilon-1)}, &
V_9 &= \frac{2(a+(1-2\epsilon)\rho)}{(a-\rho)^2(\epsilon-1)(a-\rho\epsilon)(a+(1-\epsilon))}, \\
V_{10} &= \frac{2}{(a-\rho)^2(2-\epsilon)(1-\epsilon)}, &
V_{11} &= \frac{2}{(a-\rho)^2(2a-\rho\epsilon)(a-\rho\epsilon)}.
\end{align*}

\section{Proof of the Theorem~\ref{theorem1}}
\label{appB}

For simplicity, we consider only the case $N=1$, the proof generalizes immediately to $N>1$ at the price of cumbersome notations. 
We also consider the case when the process $\vec{x}$ always belongs to $[0,1]$. 
Thus, we may assume that $V_s$, $s\in S$ and $H$ are bounded as well as their derivatives. 
We first consider the case of a Lipschitz function $\varphi$.

We denote by $(\vec{x},s)$ the PDMP starting with the random initial data distributed along $\mu_0$ and by $(\vec{x}(t,\zeta_0),s(t,\zeta_0))$ the PDMP starting from the deterministic initial data $\zeta_0=({x}_0,s_0)$.

In this proof, we use the letter $\kappa$ for a constant which depends on $V_s$, $s\in S$, $H$ and $\tau$.


The push-forward algorithm can be rewritten as follows: starting from $\tilde{s}(0) = s_0$, $\tilde{x}(0)=\vec{x}_0$, we set $\tilde{x}(\tau_{j+1}) = \vec{\Phi}_{s(\tau_j)}(\tau_{j+1}-\tau_j,\tilde{x}(\tau_j))$ and then choose $s(\tau_{j+1})$ randomly, it is equal to $r\in S$ with probability $\Pi_{s(\tau_j), r}(\tau_j,\tau_{j+1})$, the latter matrix being defined in \eqref{PI}.


In fact, we consider more generally the approximation of $\mathbb{E}(\varphi(\vec{x}(\tau),s(\tau))$, for a function $\varphi\, : \, \R\times S\to\R$, by $\mathbb{E}(\varphi(\tilde{x}(\tau),\tilde{s}(\tau))$.
The result is obtained when we choose $\varphi$ independent on $s\in S$.

We first assume that $\varphi$ is $C^1$ with respect to $x$. 
Let us denote by $L$ its Lipschitz constant with respect to $x$:
\[
|\varphi(x_1,s)-\varphi(x_2,s)|\le L |x_1-x_2|
\]
for $x_1,\, x_2 \in \R$, $s\in S$. 

The main tool for the proof is the generator of the process:
\[
\mathcal L \varphi(x,s)=V_s(x)\partial_x \varphi +\sum_{r\in S}(\varphi(x,r)-\varphi(x,s))H_{r,s}(x),
\]
and the forward Kolmogorov equation:
\[
\frac{du}{dt}(t,x,s)=\mathcal L u(t,x,s), \; x\in \R,\, s\in S,\, t> 0.
\]
When $u(0,x,s)=\varphi(x,s)$ for $x\in\R,\, s\in S$, it is well known that $u(t,x_0,s_0)=\mathbb{E}(\varphi(\vec{x}(t,\zeta_0),s(t,\zeta_0)))$ (see \cite{davis}).
Therefore,
\[
\begin{split}
\mathbb{E}(\varphi(\vec{x}(t),s(t)) & = \int_{\R\times S} \mathbb{E}(\varphi(\vec{x}(t,\zeta_0),s(t,\zeta_0)))d\mu_0(\zeta_0) \\
& = \int_{\R\times S} u(t,x_0,s_0)d\mu_0(x_0,s_0).
\end{split}
\]
 
It follows from Proposition A.1 in \cite{crudu2011convergence} that $u$ satisfies the formula:
\[
\begin{split}
u & (t,x,s)  =\varphi(\Phi_s(t,{x}),s)\exp\left(-\int_0^t \lambda(\Phi_s(\sigma,{x}))d\sigma\right) \\
& +\sum_{s\in S} \int_0^t u(t-\sigma,\Phi_s(\sigma,x),r,)H_{r,s}(\Phi_s(\sigma,x)) \exp\left(-\int_0^\sigma \lambda(\Phi_s(\tau,{x}))d\tau\right)d\sigma
\end{split}
\]
Following \cite{crudu2011convergence} we apply Gronwall's lemma to show that, when $\varphi$ is Lipschitz with respect to $x$, so is $u$, with a Lipschitz constant less than $\kappa\|\varphi\|_{0,1}$.
Here, $\kappa$ depends on the characteristics of the PDMP and the time horizon $\tau$; $\|\varphi\|_{0,1}$ is the sum of the supremum of $\varphi$ with its Lipschitz constant. 
Moreover, it follows from the forward Kolmogorov equation that its time derivative is also bounded.
Hence,
\begin{equation}\label{uLip}
|u(t_1,x_1,s)-u(t_2,x_2,s)| \le 
\kappa\|\varphi\|_{0,1}( |x_1-x_2|+|t_1-t_2|).
\end{equation}

Now we rewrite the error as follows:
\[
\begin{split}
&\mathbb{E}(\varphi(\vec{x}(\tau,\zeta_0),s(\tau,\zeta_0))- \mathbb{E}(\varphi(\tilde {x}(\tau,\zeta_0),\tilde{s}(\tau,\zeta_0)) \\[2mm]
&=\mathbb{E}(u(\tau,x_0,s_0)- u(0,\tilde x(\tau,\zeta_0), \tilde s(\tau,\zeta_0)))\\
& = \! \sum_{j=0}^{M-1} \! \mathbb{E}(u(\tau-\tau_j,\tilde x(\tau_j,\zeta_0),\tilde s(\tau_j,\zeta_0))- u(\tau-\tau_{j+1},\zeta_0,\tilde x(\tau_{j+1},\zeta_0),\tilde s(\tau_{j+1},\zeta_0))
\end{split}
\]
On each interval $[\tau_j,\tau_{j+1}]$ we have, by chain rule and the forward Kolmogorov equation, that
\begin{align*}
&\frac{d}{dt}\left[ u(\tau-t,\tilde x(t,\zeta_0), \tilde s(\tau_j,\zeta_0)) \right]\\
&=-\frac{du}{dt}(\tau-t, \tilde x(t,\zeta_0), \tilde s(t,\zeta_0))+\frac{d\tilde x}{dt}(t)\partial_x u(\tau-t,\tilde x(t,\zeta_0), \tilde s(\tau_j,\zeta_0)\\
&=-\frac{du}{dt}(\tau-t, \tilde x(t,\zeta_0), \tilde s(t,\zeta_0))+V_{\tilde s(\tau_j,\zeta_0)}(\tilde x(t,\zeta_0))\partial_x u(\tau-t,\tilde x(t,\zeta_0),\zeta_0, \tilde s(\tau_j,\zeta_0)\\
&=-\sum_{r\in S}(u(\tau-t,\tilde x(t,\zeta_0),r)-u(\tau-t,\tilde x(t,\zeta_0),\tilde s(\tau_j,\zeta_0)))H_{r,\tilde s(\tau_j,\zeta_0)}(\tilde x(t,\zeta_0)).
\end{align*}
Hence,
\[
\begin{split}
& \mathbb{E}\big(u(\tau-\tau_{j},\tilde x(\tau_{j},\zeta_0), \tilde{s}(\tau_j),\zeta_0) - u(\tau - \tau_{j+1},\tilde{x}(\tau_{j+1},\zeta_0), \tilde{s}(\tau_{j+1},\zeta_0))\big) \\
& = \mathbb{E} \left(\int_{\tau_{j}}^{\tau_{j+1}} \sum_{r\in S}(u(\tau - t,\tilde{x}(t,\zeta_0),r) - u(\tau - t,\tilde{x}(t,\zeta_0),\tilde{s}(\tau_j,\zeta_0))) H_{r,\tilde{s}(\tau_j,\zeta_0)}(\tilde x(t,\zeta_0)) dt \right) \\[2mm]
& + \mathbb{E} (u(\tau - \tau_{j+1},\tilde{x}(\tau_{j+1},\zeta_0), \tilde{s}(\tau_{j},\zeta_0)) - u(\tau-\tau_{j+1},\tilde{x}(\tau_{j+1},\zeta_0), \tilde{s}(\tau_{j+1},\zeta_0))) \\[2mm]
& = \mathbb{E} \left(\int_{\tau_{j}}^{\tau_{j+1}} \sum_{r\in S}(u(\tau-t,\tilde x(t,\zeta_0),r)-u(\tau-t,\tilde{x}(t,\zeta_0),\tilde s(\tau_j,\zeta_0)))H_{r,\tilde{s}(\tau_j,\zeta_0)}(\tilde x(t,\zeta_0)) dt \right) \\
& + \mathbb{E} \left(\sum_{r\in S}(u(\tau-\tau_{j+1},\tilde x(\tau_{j+1},\zeta_0)), \tilde{s}(\tau_j,\zeta_0)-u(\tau-\tau_{j+1},\tilde x(\tau_{j+1},\zeta_0),r) \right) \Pi_{r,\tilde{s}(\tau_,\zeta_0j)}(\tau_j,\tau_{j+1}).
\end{split}
\]
By eq. \eqref{PI}, we have that 
\[
\Pi(\tau_j,\tau_{j+1})=I+\int_{\tau_j}^{\tau_{j+1}} H(\tilde x(t),\zeta_0)\Pi(\tau_j,t)dt,
\]
where $I$ is the identity matrix. 

Since $H$ is bounded, it follows from Gronwall's lemma that $\Pi$ is bounded and 
$$
\| \Pi(\tau_j,t)-I\| \le \kappa |t-\tau_j|
$$
for some constant $\kappa$, where $\|\cdot\|$ is a norm on the space of matrices. 
For instance we may take the supremum of the modulus of the coefficients of a matrix.
Hence,
\begin{align*}
\|\Pi(\tau_j,\tau_{j+1})-I -\int_{\tau_j}^{\tau_{j+1}} H(\tilde x(t,\zeta_0))dt\|&= \| \int_{\tau_j}^{\tau_{j+1}} H(\tilde x(t,\zeta_0))(I-\Pi(\tau_j,t))dt\|\\
&\le \kappa |\tau_{j+1}-\tau_j|^2,
\end{align*}
for another constant $\kappa$. 
In particular, for $r\ne \tilde s(\tau_j)$:
\[
|\Pi_{r,\tilde s(\tau_j)}(\tau_j,\tau_{j+1}) -\int_{\tau_j}^{\tau_{j+1}} H_{r,\tilde s(\tau_j,\zeta_0)}(\tilde x(t,\zeta_0))dt\|
\le \kappa |\tau_{j+1}-\tau_j|^2,
\]
Moreover, 
\[
|\tilde{x}(\tau_{j+1,\zeta_0)}-\tilde x(t,\zeta_0)|
\le \kappa |\tau_{j+1}-t|,
\]
since its derivative is $V_{\tilde s_j}(\tilde x(t))$, which is bounded. 
We have shown in \eqref{uLip} that $u$ is Lipschitz with respect to $x$ and $t$, thus
\begin{equation}\label{uLip2}
|u(\tau-t,\tilde x(t,\zeta_0),r)-u(\tau-\tau_{j+1},\tilde x(\tau_{j+1},\zeta_0),r)|\le \kappa(\|\varphi\|_{0,1}+1) |\tau_{j+1}-t|.
\end{equation}

Now we have the following estimate
\begin{align*}
& |\mathbb{E}(u(\tau-\tau_{j},\tilde x(\tau_{j},\zeta_0), \tilde s(\tau_j,\zeta_0))-u(\tau-\tau_{j+1},\tilde x(\tau_{j+1},\zeta_0), \tilde s(\tau_{j+1},\zeta_0)))| \\[2mm]
& = \left|\mathbb{E} \left( \int_{\tau_{j}}^{\tau_{j+1}} \sum_{r\in S}(u(\tau-t,\tilde x(t,\zeta_0),r)-u(\tau-t,\tilde x(t,\zeta_0),\tilde s(\tau_j,\zeta_0)))H_{r,\tilde s(\tau_j,\zeta_0)}(\tilde x(t,\zeta_0))dt\right) \right| \\
& + \mathbb{E} \left(\sum_{r\in S}(u(\tau-\tau_{j+1},\tilde x(\tau_{j+1},\zeta_0), \tilde{s}(\tau_j,\zeta_0))-u(\tau-\tau_{j+1},\tilde{x}(\tau_{j+1},\zeta_0),r))\Pi_{r,\tilde s(\tau_j,\zeta_0)}(\tau_j,\tau_{j+1})\right)\\
& \le \left|\mathbb{E} \left(\int_{\tau_{j}}^{\tau_{j+1}} \sum_{r\in S}(u(\tau-t,\tilde x(t,\zeta_0),r)-u(\tau-t,\tilde x(t,\zeta_0),\tilde s(\tau_j,\zeta_0))) \right. \right. \\
& + (u(\tau-\tau_{j+1},\tilde x(\tau_{j+1},\zeta_0)), \tilde s(\tau_j,\zeta_0)-u(\tau-\tau_{j+1},\tilde x(\tau_{j+1},\zeta_0),r))H_{r,\tilde s(\tau_j),\zeta_0}(\tilde x(t,\zeta_0)) dt \Bigg) \Bigg| \\
& + \kappa |\tau_{j+1}-\tau_j|^2 \\[2mm]
& \le \kappa (\|\varphi\|_{0,1}+1)|\tau_{j+1}-\tau_j|^2
\end{align*}
and so
\[
\begin{split}
|\mathbb{E}(\varphi(\vec{x}(\tau,\zeta_0),s(\tau,\zeta_0))- \mathbb{E}(\varphi(\tilde {x}(\tau,\zeta_0),\tilde{s}(\tau,\zeta_0))|
& \le \sum_{j=0}^{M-1} \kappa |(\|\varphi\|_{0,1}+1)\tau_{j+1}-\tau_j|^2\\
& \le C\kappa\tau (\|\varphi\|_{0,1}+1)/M.
\end{split}
\]
The final error estimate is obtained by taking $\varphi$ depending only on $x$ and integrating with respect to $\mu_0$:
\[
\begin{split}
\left|\int_{[0,1]}\varphi(x) \mu_t^M(dx)\right. & - \left. \int_{B}\varphi(x) \mu_t(dx)\right| \\
& = \left|\int_{[0,1]} \mathbb{E}(\varphi(\vec{x}(\tau,\zeta_0)))- \mathbb{E}(\varphi(\tilde {x}(\tau,\zeta_0))) d\mu_0(\zeta_0)\right| \\[1mm]
& \le C\kappa\tau (\|\varphi\|_{0,1}+1)/M 
\end{split}
\]

When $\varphi$ is only Lipschitz, we obtain the same result by choosing a sequence of $C^1$ functions which converges uniformly to $\varphi$ and whose Lipschitz constants do not exceed $L$. 
This finishes the proof of the first statement. 
It is classical that this implies that the left-handed side converges to zero when 
\[
\max_{j=0,\dots,M-1} |\tau_{j+1}-\tau_j|\to 0
\]
for any uniformly continuous and bounded function $\varphi$. 
We conclude thanks to the Portemanteau theorem \cite{billingsley}.

\begin{remark} \label{r} \rm
When $\varphi$ is only borelian bounded but $\mu_0$ has a $C^1$ and Lipschitz density $f_0$, the proof can be adapted. 
We observe that the only point where the Lipschitz property of $\varphi$ is used is to obtain \eqref{uLip}  which is used to derive \eqref{uLip2}.
To replace \eqref{uLip2}, we can instead write the explicit formula for $\mathbb{E}(\varphi( \vec{x}(t),{s}(t))$ and $\mathbb{E}(\varphi(\tilde {x}(t),\tilde{s}(t))$. 
Since the flows $\vec{\Phi}_s,\; s\in S$ are $C^1$ diffeomorphisms that depend smoothly on time, we see that \eqref{uLip2} still holds, but on the left-handed side, $\kappa(\|\varphi\|_{0,1}+1)$ must be replaced by a constant which depends on $\|f_0\|_{0,1}$ and the supremum of $\varphi$.
\end{remark}

\end{document}